\documentclass{elsart}

\usepackage{bm}

\newcommand{\nn}{\nonumber}

\newcommand{\BV}{\mbox{$\langle 0 |$}}
\newcommand{\KV}{\mbox{$| 0 \rangle$}}
\newcommand{\bra}[1]{\mbox{$\langle #1|$}}
\newcommand{\ket}[1]{\mbox{$| #1\rangle$}}
\newcommand{\braket}[2]{\mbox{$\langle #1| #2\rangle $}}

\newcommand{\ep}{\varepsilon \bar{\epsilon}}
\begin{document}

\begin{frontmatter}



\title{Goldstone Theorem, Hugenholtz-Pines Theorem and Ward-Takahashi Relation in Finite
 Volume Bose-Einstein Condensed Gases}


\author[HE]{Hiroaki Enomoto},
\ead{enomotohiroaki@akane.waseda.jp}
\author[MO]{Masahiko Okumura} and
\ead{okumura@aoni.waseda.jp}
\author[YY]{Yoshiya Yamanaka}
\ead{yamanaka@waseda.jp}

\address[HE]{Department of Physics, Waseda University, Tokyo 169-8555,
 Japan}
\address[MO]{Department of Applied Physics, Waseda University, Tokyo
 169-8555, Japan}
\address[YY]{Department of Materials Science and Engineering, Waseda
 University, \\ Tokyo 169-8555, Japan}

\begin{abstract}
 We construct an approximate scheme based on the concept of the
spontaneous symmetry
 breakdown, satisfying the Goldstone theorem, for finite volume
 Bose-Einstein condensed gases in both zero and finite temperature cases.
 In this paper, we discuss the Bose-Einstein condensation in a box
 with periodic boundary condition and do not assume the thermodynamic
 limit.
 When energy spectrum is discrete, we found that it is necessary to deal
 with the Nambu-Goldstone mode explicitly without the Bogoliubov's
 prescription, in which zero-mode creation- and annihilation-operators
 are replaced with a {\it c}-number
 by hand, for satisfying the Goldstone theorem.
 Furthermore, we confirm that the unitarily inequivalence of vacua
in the spontaneous symmetry breakdown is true for the finite volume
 system.
\end{abstract}

\end{frontmatter}

\section{Introduction}

Various theoretical studies on weakly interacting dilute Bose-Einstein
condensed systems have been done over many years.
The excitation spectrum in a homogeneous system was first obtained by
Bogoliubov in 1947 \cite{Bog}. Afterwards the quantum correction in
zero-temperature case was evaluated by Hugenholtz and Pines \cite{HP}.
In course of their discussion, they showed the Hugenholtz-Pines (HP)
theorem, which relates the chemical potential with the self-energy in a
nonperturbative way.
In these works one follows the Bogoliubov's prescription \cite{Bog}, in
which the zero-mode creation- and annihilation-operators $a_{\bm 0}$ and
$a_{\bm 0}^\dagger$ are replaced with a {\it c}-number $\sqrt{N_{\rm c}}$
by hand, where $N_{\rm c}$ is the number of condensate particles, and the
{\it c}-number is regarded as the order parameter.
Though this prescription was simple and
successful in deriving the excitation spectrum, its naive use is not consistent
in quantum theory because the canonical commutation relations (CCRs) are not
respected \cite{OY}.

In modern quantum field theory, the order parameter is introduced as a
vacuum expectation value of a quantum field.
The self-consistent mechanism for creating the order parameter is
the spontaneous symmetry breakdown (SSB).
We have many examples of the SSB, appearing in ferromagnetic,
superconducting, crystal orders, and so on \cite{U}.
The scenario of the SSB reads as follows: Suppose that an action or a
Lagrangian is invariant under a continuous transformation but the vacuum
does not show the original symmetry.
Then the non-zero vacuum expectation value of the quantum field is
induced, and is identified as the order parameter. In general, when a
continuous symmetry is broken spontaneously, there must exist a gapless
mode, called the Nambu-Goldstone (NG) mode, which is necessary to keep
the original invariance of the action. This statement is the Goldstone
theorem \cite{Gthe}, and it is proven from the Ward-Takahashi (WT)
relations \cite{WT,U}. The WT relations are in turn identities, derived
from the CCRs, the Heisenberg equation of motion, and the transformation
property.

The dilute gas systems of bosonic neutral atoms at recent experiments
\cite{C,K,H}, forming  Bose-Einstein condensates (BECs),
are described by the model Lagrangian invariant under a global phase
transformation. The appearance of BECs can be interpreted as a
spontaneous breakdown of the global phase symmetry.
For such a model in homogeneous situations,
Hohenberg and Martin \cite{HM} showed that the HP theorem is a
consequence of the WT relation with respect to the global phase
transformation first, and generalized it to a finite temperature case.
Recently, Boyanovsky et~al \cite{Boy} showed that the HP theorem holds at
one-loop level in a homogeneous BEC system.

The concept of the SSB is inherent in quantum field theory but not in
quantum mechanics because it is closely related to an infinite number of
degrees of freedom, causing many unitarily inequivalent vacua.
For spatially homogeneous systems of quantum field
one finds inequivalent vacua under the thermodynamics limit,
defined as that of infinite volume while its ratio
to the number of particles is kept finite \cite{U}.

The whole story above applies in the case of homogeneous systems
and can not be extended to spatially inhomogeneous cases without
due consideration. The BEC systems of neutral atoms at recent
experiments have finite system size and spatial inhomogeneity due to trapping potentials.
For such systems of finite size,
the invariance of spatial translation is lost and the momentum is no longer
a good quantum number, and the energy spectrum including the NG mode
becomes discrete. The issue of existence of inequivalent vacua for
such finite volume systems has not been considered thoroughly, so
the occurrence of the SSB in the systems is not trivial.
Meanwhile, it has been observed in the recent experiment on trapped
Bose gases \cite{exp1} that a condensate has a fixed phase.
Furthermore, it is claimed that the Bogoliubov spectrum \cite{exp2}
and the Bogoliubov transformation \cite{exp3} were observed in the
two-photon Bragg scattering process.
Even in the case of discrete spectrum, the theoretical formulation
of quantum field with introduction of the order parameter is consistent
with the experiment \cite{exp4}.
These experiments seem to support that as a result of the SSB the order
parameter exists with a fixed phase, not only in homogeneous infinite
volume system, but also in inhomogeneous system of finite size.

When an order parameter appears, a single vacuum should be selected among
infinite possible ones and a quasi-particle picture should be
established in a manner consistent with the Goldstone theorem.
This is a vital but not trivial step in theoretical calculation.
One needs to invent a selection procedure in theory.
In the Bogoliubov prescription for a homogeneous system, the vacuum is
selected through replacing $a_{\bm 0}$ and $a_{\bm 0}^\dagger$ with
$\sqrt{N_{\rm c}}$ by hand, and the Bogoliubov transformation of ${\bm
p}\neq{\bm 0}$ modes diagonalizes the unperturbed Hamiltonian.
It is remarked that the zero-mode operators are absent in the
formulation so that the CCRs for the field operators and the Goldstone
theorem can not be satisfied exactly. But since the point of the
zero-energy is embedded in a continuous spectrum, this fact is easily
overlooked.
In case of finite size system such as the trapped BEC systems with
discrete spectrum, the Bogoliubov prescription clearly fails: The CCRs
of the quantum field and the Goldstone theorem are violated due to the
absence of the discrete zero-mode.
Conventionally, in the SSB of quantum field theory, one adopts the
systematic method to introduce an infinitesimal symmetry breaking term,
known as Bogoliubov's quasi-average \cite{Bogoqua}, and one vacuum is
selected among many.
Note that the zero-mode is naturally included in this method in the
symmetric limit. We have proposed the way of introducing the
infinitesimal symmetry breaking term for the trapped BEC system and have
explicitly shown that the WT relations and the Goldstone theorem hold at
tree level \cite{OY}.
It is also shown that the vacua belonging to the order parameters with
different phases are unitarily inequivalent to each other
even for the trapped BEC \cite{InEqOY}.

Our goal is a formulation of a systematic and consistent treatment of
inhomogeneous quantum field system of finite size, corresponding to the
recent experiments of the trapped BEC.
We mean by the word ``consistent treatment'' that the CCR, the WT
relations and the Goldstone theorem are respected.
As mentioned above, we have studied this subject at tree level
\cite{OY}, and should extend our discussions to any loop level.
At present, such extension is not easy due to the complex structure of
the unperturbed propagators.

Instead, in this paper, we consider a BEC system, without trapping
potential but confined in a box with periodic boundary condition. We keep a finite
volume size $V$.
To select a vacuum, we introduce an infinitesimal symmetry
breaking term, characterized by an infinitesimal parameter
$\varepsilon$.
What is important in our present study is that the limit $\varepsilon
\rightarrow 0$ is taken but the limit $V \rightarrow \infty$ is not,
namely that the thermodynamic limit is not considered.
We will see that inequivalent vacua emerge in the limit $\varepsilon
\rightarrow 0$ with finite $V$ (without the thermodynamic limit).  While
the energy spectrum is discrete, the Fourier (momentum) representation
is available in this model, which makes calculations of loop expansions
tractable. It is expected that a model with trapping potential mostly
shares essential theoretical features with this model.

Explicitly we show that  the HP theorem is derived from the WT relations
even in this finite volume system. The quasi-particle picture in which
the zero-mode (NG mode) is present is constructed at tree and
one-loop levels.
To extend the results at zero temperature to those at finite
one, we employ thermofield Dynamics (TFD) \cite{U,TFD} which is a real
time canonical formalism of thermal field theory with doubled degrees of
freedom.

This paper is organized as follows.
In Section.~2, we give a formulation of the model Lagrangian density with an
infinitesimal symmetry breaking term at zero temperature, and construct
the quasi-particle picture consistently with the CCRs and the Goldstone
theorem.
The order parameter is introduced without using the Bogoliubov's
prescription.
We give the quantum correction which keeps the HP theorem without
taking the infinite volume limit.
We also show that the vacuum in a broken phase is orthogonal to that
in as symmetric phase.
In Section.~3, we use TFD to extend the results in Section.~2 to finite
temperature case.
We review the TFD formalism briefly and investigate the WT relations at
each loop level similarly to the zero temperature case.
In this case, we also derive the HP theorem from the WT relations at
arbitrary loop level.
Section 4 is devoted to a summary and conclusion. 
In Appendix, we derive the relation between the vcua in the broken and symmetric phase. 

\section{Zero Temperature Case}

We formulate a field-theoretical treatment for
the BECs in a box with periodic boundary condition on the
field variables. In this section, we discuss the zero temperature case.
First we confine our discussion to tree level, and will see that
the WT relation is satisfied and that the unitarily inequivalent vacua
are realized. Next we derive the HP theorem from the WT relations at
any loop level and show explicitly that
our one-loop calculation satisfies the HP theorem.

\subsection{Model Lagrangian density and Hamiltonian}

We start with the following Lagrangian density, which describes a
weakly interacting Bose gas in a box whose volume is denoted by $V$,
\begin{equation}
{\mathcal L}=\Psi^\dag(x)\left({\rm i} \hbar\frac{\partial}{\partial
t}+\frac{\hbar^{2}}{2m}\nabla^{2}+\mu\right)\Psi(x)
-\frac{g}{2} \Psi^\dag (x) \Psi^\dag (x) \Psi (x) \Psi (x) \, .
\label{L}
\end{equation}
Here, $m$, $\mu$ and $g$ are the mass of a boson, the chemical potential
and the coupling constant, respectively, and $x$ stands for the
space-time coordinate $({\bm x},t)$.  The periodic boundary condition in
each direction is imposed on the field variables:
\begin{equation}
\left\{
\begin{array}{ccc}
\Psi(x+L, y, z,t) & = & \Psi(x, y, z,t) \\
\Psi(x, y+L, z,t) & = & \Psi(x, y, z,t) \\
\Psi(x, y, z+L,t) & = & \Psi(x, y, z,t)
\end{array}
\right. \, , \label{periodic}
\end{equation}
where $L$ being the length of one side of the cube, $V=L^3$.
It is easy to see that this Lagrangian density is invariant under the
global phase transformation:
\begin{eqnarray}
 \Psi (x) & \rightarrow & e^{{\rm i} \theta} \Psi (x) \nn \\
 \Psi^\dag (x) & \rightarrow & e^{- {\rm i} \theta} \Psi^\dag (x) \, ,
  \label{gps}
\end{eqnarray}
where $\theta$ is a real constant.
When a uniform BEC is created, the global phase symmetry is
spontaneously broken, and then the quantum field $\Psi (x)$ is divided
into a classical constant field $v$ and a quantum field $\varphi (x)$ in
the terminology of the canonical operator formalism:
\begin{equation}
 \Psi (x) = v + \varphi (x) \, .
\end{equation}
We take a real value for $v$, which does not affect generality of
the following discussions.
The classical field $v$ is called the order parameter and is also expressed as
$v^2 = n_{\rm c}$ in terms of the density of condensate particles $n_{\rm c}$.
The classical field $v$ may be defined as an expectation value of the
Heisenberg field with respect to the vacuum $| \Omega \rangle$,
\begin{eqnarray}
\bra{\Omega} \Psi (x) \ket{\Omega} = v \, ,
\end{eqnarray}
or equivalently
\begin{equation}
\bra{\Omega}  \varphi (x) \ket{\Omega} = 0 \, . \label{expvphi0}
\end{equation}

Let us introduce an artificial symmetry breaking term
\begin{eqnarray}
{\mathcal L}_{\varepsilon}=(\ep) v \left[\Psi(x)+\Psi^\dag (x)\right] \, ,
\label{isbt}
\end{eqnarray}
and add it to the original Lagrangian density (\ref{L}),
then the total Lagrangian density ${\mathcal L}_{\rm tot}$ becomes
\begin{equation}
 {\mathcal L}_{\rm tot} = {\mathcal L} + {\mathcal L}_\varepsilon \, .
  \label{Ltotdef}
\end{equation}
Here, $\varepsilon$ is an infinitesimal dimensionless parameter and
$\bar\epsilon$ represents a typical energy scale of the system.
Obviously the total Lagrangian ${\mathcal L}_{\rm tot}$ is not invariant under
the global phase transformation (\ref{gps}).  The parameter $\varepsilon$
is taken to be vanishing at the final stage of the calculation, so that
the original symmetry is restored.

We move to the canonical formalism in the interaction representation.
The canonical conjugate of $\varphi (x)$ is $\pi (x) = {\rm i} \hbar \varphi^\dag
(x)$.
Then the CCRs are as follows:
\begin{equation}
[ \varphi ({\bm x}, t) , \varphi^\dag ({\bm x'}, t) ] = \delta^ 3 ({\bm x} -
 {\bm x'}) \, , \label{CCRvphi1}
\end{equation}
\begin{equation}
[\varphi ({\bm x}, t) , \varphi ({\bm x'}, t)] = [\varphi^\dag ({\bm x},
 t), \varphi^\dag ({\bm x'}, t)] = 0 \, . \label{CCRvphi2}
\end{equation}
Now, we have the total Hamiltonian
\begin{equation}
H = H_0 + H_{\rm{int}} + {\rm{const.}} \, ,
\end{equation}
where the unperturbed Hamiltonian $H_0$ is given as
\begin{eqnarray}
H_0 & = & \int \! {\rm d}^3 x \, \Bigg[ \varphi^\dag (x) \left( -
\frac{\hbar^{2}}{2m} \nabla^2 - \mu \right) \varphi (x) \nn \\
& & + \frac{gv^2}{2}
\left\{ 4 \varphi^\dag (x) \varphi (x) + \varphi (x) \varphi (x) +
\varphi^\dag (x) \varphi^\dag (x) \right\} \Bigg] \, . \label{H0vphi}
\end{eqnarray}
and the perturbative Hamiltonian $H_{\rm int}$ is defined as
\begin{eqnarray}
H_{\rm{int}} & = & v \left( - \mu + gv^2 - \ep \right)\int \! {\rm d}^3 x \,
 \left\{ \varphi (x) + \varphi^\dag (x) \right\} \nn \\
& & {} + gv \int \! {\rm d}^3 x \, \left[ \varphi^\dag (x) \varphi^\dag (x)
\varphi (x) + \varphi^\dag (x) \varphi (x) \varphi (x) \right] \nn \\
& & {} +\frac{g}{2}\int \! {\rm d}^3 x \,  \varphi^\dag (x) \varphi^\dag (x)
\varphi (x) \varphi (x) \, .
\label{Hintvphi}
\end{eqnarray}

It is convenient to define the 2$\times$2-matrix propagator for the
quantum fields $\varphi (x)$ and its Hermite conjugate $\varphi^\dag (x)$
by
\begin{eqnarray}
G(x-x') & = &
\left(
\begin{array}{cc}
G_{11} (x - x') & G_{12} (x - x') \\
G_{21} (x - x') & G_{22} (x - x')
\end{array}
\right) \nn \\
& = &
\left(
\begin{array}{cc}
- {\rm i} \bra{\Omega} {\bf {\rm T}} [\varphi (x) \varphi^\dag (x') ] \ket{\Omega} &
- {\rm i} \bra{\Omega} {\bf {\rm T}} [\varphi (x) \varphi (x') ] \ket{\Omega} \\
- {\rm i} \bra{\Omega} {\bf {\rm T}} [\varphi^\dag (x) \varphi^\dag (x') ] \ket{\Omega} &
- {\rm i} \bra{\Omega} {\bf {\rm T}} [\varphi^\dag (x) \varphi (x') ] \ket{\Omega}
\end{array}
\right) \, ,
\end{eqnarray}
where ${\bf {\rm T}}$ is the symbol for time-ordered product.
Its Fourier transform is given by
\begin{equation}
 G (p) =
\left(
\begin{array}{cc}
 G_{11} (p) & G_{12} (p) \\
 G_{21} (p) & G_{22} (p)
\end{array}
\right)
= \int \! \frac{{\rm d}^4 x}{(2 \pi \hbar)^2} G (x )
  e^{\frac{{\rm i}}{\hbar}({\bm p} \cdot {\bm x} - \omega t)} \, ,
\end{equation}
with the notation of $p=(\omega, {\bm p})$.
It is mentioned that one has the following relations among
the matrix elements:
\begin{eqnarray}
G_{11} (p) = G_{22} (-p) \, , \quad  G_{12} (p) = G_{21} (p) \, .
 \label{relGp}
\end{eqnarray}

\subsection{Derivation of the Ward--Takahashi relation}

It is well-known that the Goldstone theorem follows from the
WT relations. We adopt the loop expansion, because the WT relations hold
at each loop level in general.

First we review the WT relation in general case.
Consider an infinitesimal transformation:
\begin{equation}
 \Psi (x) \rightarrow \Psi' (x) = \Psi (x) + \xi \delta \Psi (x)
\end{equation}
with the infinitesimal parameter $\xi$.
The change in the Lagrangian density is denoted by $\xi \delta {\mathcal
L}$:
\begin{equation}
 \xi \delta {\mathcal L} = {\mathcal L} [\Psi' (x)] - {\mathcal L} [ \Psi (x) ] \, .
  \label{delLdef}
\end{equation}
The N{\" o}ther theorem implies
\begin{equation}
 \delta {\mathcal L} = \partial^\mu N_\mu \, , \label{NT}
\end{equation}
where
\begin{equation}
 N_\mu = \frac{\partial {\mathcal L}}{\partial \Psi^\mu} \delta \Psi (x) \, .
\end{equation}
Now, we define the N\"other charge
\begin{equation}
 N (t) = \frac{1}{\hbar}\int \! {\rm d}^3 x \, N_0 (x) \, ,
\end{equation}
which generates the transformation in the commutation relation
\begin{equation}
 [ \Psi (x), N(t) ]_{t_x =t} = {\rm i} \delta \Psi (x) \, .
\end{equation}
Then, the N{\" o}ther theorem (\ref{NT}) leads to
\begin{equation}
 \dot{N} (t) = \frac{1}{\hbar} \int \! d^3 x \, \delta {\mathcal L} \, .
\end{equation}
One can easily derive the following relation,
\begin{eqnarray}
& & \frac{\partial}{\partial t} \bra{\Omega} {\bf {\rm T}} [ N (t) \Psi
 (x_1) \cdots \Psi (x_n) ] \ket{\Omega} \nn \\
& = & \sum_{a = 1}^{n} \delta (t - t_a) \bra{\Omega} {\bf {\rm T}}
 [ \Psi (x_1) \cdots [ N (t), \Psi (x_a) ] \cdots \Psi (x_n) ]
 \ket{\Omega} {} \nn \\
& & {} + \bra{\Omega}{\bf {\rm T}} [ \dot{N} (t) \Psi (x_1) \cdots \Psi
 (x_n) ] \ket{\Omega} \, .
\end{eqnarray}
Integrating both sides over the entire time domain and dropping the
surface term at $t = \pm \infty$, we obtain
\begin{equation}
 \sum_{a = 1}^{n} {\rm i} \hbar \bra{\Omega} {\bf {\rm T}} [\Psi (x_1) \cdots
  \delta \Psi (x_a) \cdots \Psi (x_n) ] \ket{\Omega} = \int \! {\rm d}^3 x
  \bra{\Omega} {\bf {\rm T}} [ \delta {\mathcal L} (x) \Psi (x_1) \cdots
  \Psi (x_n) ] \ket{\Omega} \, .
 \label{WTgeneral}
\end{equation}
This relation, derived independently of any approximate scheme, is
called the WT relation.

In this paper, we are interested in the following infinitesimal global
phase transformation:
\begin{equation}
 \delta \Psi (x) = {\rm i} \Psi (x) \, .
\end{equation}
The N{\" o}ther charge is given by
\begin{equation}
 N(t) = -  \int \! {\rm d}^3 x \, \Psi^\dag (x) \Psi (x) \, ,
\end{equation}
indeed $[N(t),\Psi(x)]_{t_x=t}=\Psi(x)=-{\rm i} \delta\Psi(x)$.
We consider the following special form of the WT relation
(\ref{WTgeneral}):
\begin{eqnarray}
{\rm i} \hbar \langle \Omega | \delta \delta {\mathcal L}_{\rm tot} (x) | \Omega
\rangle = \int \! {\rm d}^4 x'\, \langle \Omega | {\rm T} [ \delta {\mathcal L}_{\rm
tot} (x') \delta {\mathcal L}_{\rm tot} (x) ] | \Omega \rangle \, .
\end{eqnarray}
Here, $\delta {\mathcal L}_{\rm tot} (x)$ and $\delta \delta {\mathcal L}_{\rm
tot} (x)$ are defined as
\begin{equation}
\delta {\mathcal L}_{\rm tot} (x) = \delta {\mathcal L}_{\varepsilon} (x) =
 {\rm i} [N(t),{\mathcal L}_{\varepsilon} (x) ]_{t_x=t} = {\rm i} (\ep) v \left[ \Psi (x) -
 \Psi^\dag (x) \right]
\end{equation}
and
\begin{equation}
\delta \delta {\mathcal L}_{\rm tot} = \delta \delta {\mathcal L}_{\varepsilon}
 = {\rm i} [ N(t), \delta {\mathcal L}_{\varepsilon} (x) ]_{t_x=t} = - (\ep) v \left[ \Psi
(x) + \Psi^\dag (x) \right] \, .
\end{equation}
Thus, the WT relation is written in the form of
\begin{eqnarray}
v & = & \frac{-(\ep)v}{2\hbar} \int \! {\rm d}^4 x' \, \left[ G_{11} (x-x') +
G_{22} (x-x') - G_{12} (x-x') - G_{21} (x-x') \right] \nn \\
& = & - \frac{(\ep)v}{\hbar} \left[ G_{11} (p=0) - G_{12} (p=0) \right] \,
 , \label{WT1}
\end{eqnarray}
where we have used the relations (\ref{relGp}) in the last line.

\subsection{Diagonalization of unperturbed Hamiltonian and the 
Ward--Takahashi relation at tree level\label{DH0}}

In this subsection, we investigate the system at tree level.
First let us determine the order parameter at tree level.
Form the unperturbed Hamiltonian (\ref{H0vphi}) and the first term in
the perturbative Hamiltonian (\ref{Hintvphi}), we have the diagram at
tree (zero-loop) level for the condition (\ref{expvphi0}), corresponding
to the expression
\begin{equation}
\left(
\begin{array}{c}
 \bra{\Omega} \varphi (x) \ket{\Omega} \\
 \bra{\Omega} \varphi^\dag (x) \ket{\Omega}
\end{array}
\right)
= {\rm i} \int \! {\rm d}^4 x' \,  G (x-x') (- \mu_0 + gv^2 - \ep) = 0 \, , \label{tad}
\end{equation}
where $\mu_0$ is the chemical potential at tree level.
As this equation must be true for any $x$, $\mu_0$ is determined as
\begin{equation}
 \mu_0 = g v^2 - \ep \, . \label{mugv2}
\end{equation}

Next, we construct the quasi-particle picture at tree level by
diagonalizing the unperturbed Hamiltonian (\ref{H0vphi}).
The unperturbed quantum field $\varphi (x)$ is expanded in terms of
annihilation-operators $a_{\bm p} (t)$ as
\begin{eqnarray}
\varphi (x) =\frac{1}{\sqrt{V}}  \sum_{{\bm p}={\bm - \infty}}^{\infty}
a_{\bm p}(t) e^{\frac{{\rm i}}{\hbar} {\bm p} \cdot
{\bm x}} \, . \label{vphia}
\end{eqnarray}
Here, $V$ is finite and the periodic boundary conditions on field operators
(\ref{periodic}) restrict ${\bm p}$ as
\begin{equation}
{\bm p} = \frac{2\pi \hbar}{L} (n_x,n_y,n_z),  \qquad\qquad
\mbox{$n_x,n_y,n_z$: integers} \, .
\end{equation}
The symbol $\sum_{{\bm p} = - \infty}^\infty$ stands for
$\sum_{n_x,n_y,n_z = - \infty}^{\infty}$ which includes $n_x = n_y = n_z
= 0$.
We emphasize that if one employs the Bogoliubov's prescription, the
quantum field does not contain $a_{\bm 0}(t)$ and breaks the CCRs,
but our expansion contains it and the CCRs are held.
The operators $a_{\bm p}(t)$ and $a^\dag_{\bm p}(t)$ are subject to
the commutation relations of creation- and annihilation-operators, which
follow from the CCRs for quantum fields (\ref{CCRvphi1}) and
(\ref{CCRvphi2}): $[a_{\bm p}(t), a^\dag_{\bm p'}(t) ] = \delta_{{\bm
p}{\bm p'}}$ and other commutations vanish.
The temporal development of $\varphi (x)$ and $a_{\bm p}(t)$ is
generated by the unperturbed Hamiltonian (\ref{H0vphi}).
Substituting the expansion (\ref{vphia}) and its Hermite conjugate into
the unperturbed Hamiltonian (\ref{H0vphi}), we obtain
\begin{equation}
H_0 = \sum_{{\bm p}= - \infty}^\infty \left[ \left( \epsilon_{\bm p} + g
v^2 + \ep \right) a^\dag_{\bm p} a_{\bm p} + \frac{gv^2}{2} \left(
a_{\bm p} a_{-{\bm p}} + a^\dag_{\bm p} a^\dag_{-{\bm p}} \right)
\right] \, , \label{H0a}
\end{equation}
where we use the relation (\ref{mugv2}) and the notation of
\begin{equation}
 \epsilon_{\bm p} = \frac{{\bm p}^2}{2m} \, .
\end{equation}
As is well known, the Bogoliubov transformation
\begin{eqnarray}
a_{\bm p}=u_{\bm p} b_{\bm p} + v_{\bm p} b^\dag_{-{\bm p}} \, ,
 \label{BTtree}
\end{eqnarray}
where
\begin{eqnarray}
u_{\bm p}, v_{\bm p} = \pm~\sqrt{\frac{\epsilon_{\bm p} + g v^2 +
\ep}{2E_{\bm p}} \pm \frac{1}{2}}
\end{eqnarray}
and
\begin{equation}
E_{\bm p} = \sqrt{\epsilon_{\bm p}^2 + 2 g v^2 \epsilon_{\bm p} + 2
\left( \epsilon_{\bm p} + g v^2 \right) \ep + (\ep)^2} \, , \label{Ep}
\end{equation}
diagonalizes the unperturbative Hamiltonian as
\begin{eqnarray}
H_0= \sum_{{\bm p}= - \infty}^\infty E_{\bm p} b^\dag_{\bm p} b_{\bm p}
 + {\rm const.}
\end{eqnarray}
We mention again that a crucial point here is that the mode with ${\bm
p}={\bm 0}$ is included in the transformation.
The infinitesimal symmetry breaking term (\ref{isbt}) makes the
transformation between $a_{\bm 0}$ and $b_{\bm 0}$ well-defined as long
as $\varepsilon$ is kept finite.

The field operator $\varphi (x)$ is rewritten in terms of the
creation- and annihilation-operators $b_{\bm p}$ and $b^\dag_{\bm p}$ as
\begin{eqnarray}
\varphi (x) = \frac{1}{\sqrt{V}}\sum_{{\bm p}= - \infty}^\infty
\left[ b_{\bm p} u_{\bm p}e^{\frac{{\rm i}}{\hbar}({\bm p} \cdot {\bm x} - E_{\bm
 p} t)} +b^\dag_{\bm p} v_{\bm p} e^{-\frac{{\rm i}}{\hbar} ({\bm p}\cdot{\bm
 x} - E_{\bm p} t)} \right] \, .
\end{eqnarray}
One can easily check that this quantum field and its Hermite conjugate
satisfy the CCRs (\ref{CCRvphi1}) and (\ref{CCRvphi2}).

The unperturbed matrix propagator of the field $\varphi(x)$ on the
vacuum $\ket{\Omega_0}$, which is defined by the relation $b_{\bm
p}\ket{\Omega_0}=0$, is given by
\begin{eqnarray}
G_0 (x-x') & =&
\left(
\begin{array}{cc}
G_{0,11} (x - x') & G_{0,12} (x - x') \\
G_{0,21} (x - x') & G_{0,22} (x - x')
\end{array}
\right) \nn \\
& = &
\left(
\begin{array}{cc}
-{\rm i} \bra{\Omega_0} {\bf {\rm T}} [ {\varphi} (x)
 {\varphi}^\dag (x') ] \ket{\Omega_0} &
-{\rm i} \bra{\Omega_0} {\bf {\rm T}} [ {\varphi} (x)
{\varphi} (x') ] \ket{\Omega_0} \\
-{\rm i} \bra{\Omega_0} {\bf {\rm T}} [ {\varphi}^\dag (x)
 {\varphi}^\dag (x') ] \ket{\Omega_0} &
-{\rm i} \bra{\Omega_0} {\bf {\rm T}} [ {\varphi}^\dag (x)
 {\varphi} (x') ] \ket{\Omega_0}
\end{array}
\right) \, .
\end{eqnarray}
Its Fourier transform is defined by
\begin{equation}
G_0 (p) =
\left(
\begin{array}{cc}
 G_{0,11} (p) & G_{0,12} (p) \\
 G_{0,21} (p) & G_{0,22} (p)
\end{array}
\right)
= \int \! \frac{{\rm d}^4 x}{(2 \pi \hbar)^2} G_0 (x )
e^{\frac{{\rm i}}{\hbar}({\bm p} \cdot {\bm x} - \omega t)} \, ,
\end{equation}
and the explicit forms of its matrix elements are
\begin{eqnarray}
G_{0,11} (p) & = & G_{0,22} (-p)
 = \frac{u_{\bm p}^2}{\omega - \omega_{\bm p} + {\rm i} \delta} - \frac{v_{\bm
 p}^2}{\omega + \omega_{\bm p} - {\rm i} \delta} \, ,  \label{propp1} \\
G_{0,12} (p) & = & G_{0,21} (p)
= \frac{u_{\bm p} v_{\bm p}}{\omega - \omega_{\bm p} + {\rm i} \delta} -
\frac{u_{\bm p} v_{\bm p}}{\omega + \omega_{\bm p} - {\rm i} \delta} \, ,
\label{propp2}
\end{eqnarray}
where $\hbar \omega_{\bm p} = E_{\bm p}$ and $\delta$ is an
infinitesimal positive parameter.

Next, let us check the WT relation (\ref{WT1}) at tree level, using
the propagators (\ref{propp1}) and (\ref{propp2}). The right-hand side
(RHS) of the relation (\ref{WT1}) is manipulated as
\begin{eqnarray}
-\frac{(\ep)v}{\hbar} \left[ G_{0,11} (p=0) - G_{0,12} (p=0) \right] & = &
-\frac{(\ep)v}{\hbar} \left(\frac{-1}{\omega_{{\bm p}={\bm 0}}}\right)
\left( \frac{g v^2 + \ep}{\hbar \omega_{{\bm p}={\bm 0}}} +
 \frac{gv^2}{\hbar \omega_{{\bm p}={\bm 0}}} \right) \nn \\
& = & \frac{2gv^3 (\ep) + (\ep)^2v}{\hbar^2 \omega_{{\bm p}={\bm 0}}^2}
 \nn \\
& = & v \, ,
\end{eqnarray}
where the last equality comes from the quasi-particle energy (\ref{Ep}).
This way the WT relation (\ref{WT1}) at tree level is confirmed.
We observe that the contribution of the zero-energy mode (the NG mode)
is vital.
In other words, for a finite volume system with a discrete spectrum, the
Bogoliubov's prescription without operators representing the NG mode can not
preserve the WT relations.

\subsection{Unitarily inequivalent vacua}

In the previous subsection, we considered the vacuum $\ket{\Omega_0}$
associated with the operator $b_{\bm p}$ as a physical one at tree
level.
There is another vacuum, denoted by $\KV$, which is annihilated by
$a_{\bm p}$: $a_{\bm p}\KV=0$.
We evaluate the inner product $\BV \Omega_0\rangle$ below, in the limit
$\varepsilon\rightarrow0$ but keeping finite $V$.

For this purpose, we will give explicit transformation to relate the
two vacua, $|\Omega_0\rangle$ and $|0\rangle$, in Appendix \ref{AppA}.
We recapitulate Eq. (\ref{S0}) which is the conclution of Appendix \ref{AppA}:
\begin{eqnarray}
|\Omega_0\rangle=\frac{1}{\sqrt{u_{\bm 0}}} \exp \left[-\frac{1}{2}{\sum_{{\bm p} = -
\infty}^\infty}' \ln u_{\bm p}\right] \exp\left[ \frac{1}{2}\sum_{{\bm p}=-
\infty}^\infty \frac{v_{\bm p}}{u_{\bm p}}a^\dag_{\bm p}a^\dag_{-{\bm
p}}\right]|0\rangle \, ,
\end{eqnarray}
where the symbol ${\sum_{{\bm p} = - \infty}^{\infty}}'$ means summation
without ${\bm p} = {\bm 0}$.
Under the limit $\varepsilon \rightarrow 0$, $u_{\bm p}$ and $v_{\bm p}$
with ${\bm p} \neq {\bm 0}$ are finite, $v_{\bm 0}/u_{\bm 0}$
becomes $1$ and $u_{\bm 0}$ is divergent as
$\varepsilon^{-\frac{1}{4}}$.
Then one finds that
\begin{equation}
\langle 0|\Omega_0\rangle \sim \varepsilon^{\frac{1}{8}} \rightarrow 0
  \qquad  (\varepsilon\rightarrow0) \, .
\end{equation}

A conclusion in this subsection is that the vacua $\ket{\Omega_0}$ and
$\KV$ are orthogonal to each other in the limit $\varepsilon \rightarrow
0$ even for finite volume system.
This means that the Fock space built on $\KV$ is unitarily inequivalent
to one built on $\ket{\Omega_0}$: a state $b_{\bm p}^{\dag n} \ket{\Omega_0}$
is never obtained by the superposition of $a_{\bm p}^{\dag n'} \KV $, where
$n$ and $n'$ are integers.
We emphasize that the existence of inequivalent representations
does not necessarily require the thermodynamic (infinite volume) limit
in this model.
This clearly shows that the choice of the unperturbed Hamiltonian and the
associated vacuum is
essential for the determination of physical quantities even in a finite
Bose--Einstein condensed systems.
For example, one can  get the Bogoliubov spectrum only for the vacuum
in a broken phase $\ket{\Omega_0}$, but never for that
in a symmetric phase $\KV$.
The fact that the Bogoliubov spectrum and the Bogoliubov transformation
are observed \cite{exp2,exp3} forces us to choose the Bogoliubov vacuum
$\ket{\Omega_0}$.

\subsection{The Hugenholtz-Pines theorem at zero temperature\label{HPTZT}}

In this subsection, we derive the HP theorem from the WT relation
at each loop level.
First, we define $\delta \mu$ as the quantum correction to $\mu_0$ (the chemical
potential at tree level),
\begin{eqnarray}
\mu = \mu_0 + \delta \mu  = g v^2 - \ep + \delta \mu \, ,
 \label{mudelmu}
\end{eqnarray}
and $\delta \mu $ will be determined later. We define the unperturbed Hamiltonian with $\mu_0$,
\begin{eqnarray}
H_0 & = & \int \! {\rm d}^3 x \, \Bigg[ \varphi^\dag (x) \left( -
\frac{\hbar^{2}}{2m} \nabla^2 - \mu_0 \right) \varphi (x) \nn \\
& & + \frac{gv^2}{2}
\left\{ 4 \varphi^\dag (x) \varphi (x) + \varphi (x) \varphi (x) +
\varphi^\dag (x) \varphi^\dag (x) \right\} \Bigg] \, , \label{H0vphi2}
\end{eqnarray}
while the perturbative Hamiltonian $H_{\rm int}$ includes the $\delta \mu$ term,
\begin{eqnarray}
H_{\rm{int}} & = & v \left( - \mu_0-\delta\mu + gv^2 - \ep \right)\int \! {\rm d}^3 x \, 
 \left\{ \varphi (x) + \varphi^\dag (x) \right\} \nn \\  
& & {} + gv \int \! {\rm d}^3 x \, \left[ \varphi^\dag (x) \varphi^\dag (x)
\varphi (x) + \varphi^\dag (x) \varphi (x) \varphi (x) \right] \nn \\
& & {} +\frac{g}{2}\int \! {\rm d}^3 x \,  \varphi^\dag (x) \varphi^\dag (x) 
\varphi (x) \varphi (x) \nn \\
& & {} - \int \! {\rm d}^3 x \, \delta\mu \varphi^\dag (x) \varphi (x) \, . 
\label{Hintvphi2}
\end{eqnarray}
Since the unperturbed Hamiltonian $H_0$ is not changed from that at tree level, 
one can adopt the quasi-particle picture and the unperturbed propagator at tree level,
given in Subsection 2.3.

Consider the Schwinger-Dyson equation,
\begin{eqnarray}
G^{-1} (p) = G_0^{-1} (p) - \Sigma (p) + \frac{\delta \mu}{\hbar} I \, ,
\end{eqnarray}
where $G(p)$ and $\Sigma (p)$ are the full propagator and the
self-energy from the loop diagrams, respectively, and $I$ is a unity $2 \times 2$ matrix.
Both of $G(p)$ and $\Sigma (p)$ are matrices and can be calculated with the elements of the
unperturbed matrix  propagator (\ref{propp1}) and (\ref{propp2}) and the perturbative
Hamiltonian (\ref{Hintvphi2}).
The matrix elements of $\Sigma (p)$ have the properties of
\begin{eqnarray}
\Sigma_{11} (p) = \Sigma_{22} (-p), \quad \Sigma_{12} (p) =
\Sigma_{21}(p) \, ,
\end{eqnarray}
where the matrix form of the self-energy is written as
\begin{equation}
 \Sigma (p) =
\left(
\begin{array}{cc}
 \Sigma_{11} (p) & \Sigma_{12} (p) \\
 \Sigma_{21} (p) & \Sigma_{22} (p)
\end{array}
\right) \, .
\end{equation}

Let us rewrite the RHS of the WT relation (\ref{WT1}) at any loop level,
we obtain that
\begin{eqnarray}
& & - \frac{(\ep) v}{\hbar} \left[ G_{11} (p=0) - G_{12} (p=0) \right] \nn \\
&=& - \frac{(\ep) v}{\hbar} \frac{G_{0,11}^{-1} (p=0) + G_{0,12}^{-1}
 (p=0) - \Sigma_{11} (p=0) - \Sigma_{12} (p=0) + \frac{\delta \mu}{\hbar}}{|G^{-1} (p=0)|} \, ,
 \label{WTloop}
\end{eqnarray}
where $|G^{-1} (p)|$ represents the determinant of $G^{-1} (p)$.
Since ${G}_{0}$ has the form in Eqs. (\ref{propp1}) and (\ref{propp2}),
one can obtain the
explicit form of ${G}_0^{-1} (p)$ as
\begin{eqnarray}
& &G^{-1}_0 (p) \nn \\
&=& \left(
\begin{array}{cc}
( \omega - \omega_{\bm p} ) u_{\bm p}^{2} - ( \omega +\omega_{\bm p}
 ) v_{\bm p}^{2} &
( \omega + \omega_{\bm p} ) u_{\bm p} v_{\bm p} - ( \omega -
\omega_{\bm p} ) u_{\bm p} v_{\bm p} \\
( \omega + \omega_{\bm p} ) u_{\bm p} v_{\bm p} - ( \omega -
 \omega_{\bm p} ) u_{\bm p} v_{\bm p} &
( \omega - \omega_{\bm p}) v_{\bm p}^{2} - ( \omega + \omega_{\bm p})
u_{\bm p}^{2} \\
\end{array}
\right) \, .
\end{eqnarray}
The determinant of $G^{-1} (p)$ is calculated as
\begin{eqnarray}
& &\left|G^{-1} (p) \right| \nn \\
& = & \left[ \omega - \omega_{\bm p} ( u_{\bm p}^{2} + v_{\bm p}^{2} ) -
 \Sigma_{11} (p) + \frac{\delta \mu}{\hbar}
\right] \left[ - \omega - \omega_{\bm p} ( u_{\bm p}^{2} + v_{\bm
p}^{2} ) - \Sigma_{22} (p) + \frac{\delta \mu}{\hbar} \right] \nn \\
& & - \left[ 2 \omega_{\bm p} u_{\bm
p} v_{\bm p} - \Sigma_{12} (p) \right]^2 \nn \\
& \simeq & - \omega^2 + \left[ \Sigma_{11} (p) - \Sigma_{22} (p) \right]
 \omega +\epsilon_{\bm p}^2 + 2gv^2 \epsilon_{\bm p} \nn \\
& & + \left( \frac{\epsilon_{\bm p}}{\hbar} + \frac{gv^2}{\hbar} \right) \left[ -
\frac{2 \delta \mu}{\hbar} + \Sigma_{11} (p) + \Sigma_{22} (p) \right]
- \frac{2 g v^2}{\hbar} \Sigma_{12} (p) \nn \\
& &+ \frac{\ep}{\hbar} \left[
\frac{2gv^2}{\hbar} - \frac{2\delta \mu}{\hbar} + \Sigma_{11} (p) + \Sigma_{22} (p) \right]
+ \frac{(\ep)^2}{\hbar^2} \, . \label{Ginv}
\end{eqnarray}
We dropped the quadratic terms of the self-energy and $\delta\mu$ in the
second line of Eq. (\ref{Ginv}), because they are quantities of
higher order.
This way the RHS of the WT relation (\ref{WT1}) at any loop level is
organized as
\begin{eqnarray}
& & - \frac{(\ep)v}{\hbar} \left[ G_{11} (p=0) - G_{12} (p=0) \right]
 \nn \\
& = & \left[\frac{(\ep)v}{\hbar} \left\{ \frac{2gv^2}{\hbar} -
 \frac{\delta \mu}{\hbar} + \Sigma_{11} (p=0) + \Sigma_{12} (p=0)
 \right\} + \frac{(\ep)^2v}{\hbar^2} \right] \nn \\
& &\times\left[\frac{2 g v^2}{\hbar} \left\{ -
 \frac{\delta \mu}{\hbar} + \Sigma_{11} (p=0) - \Sigma_{12} (p=0)
\right\} \right. \nn \\
& &\left. + 2 \frac{\ep}{\hbar} \left\{ \frac{gv^2}{\hbar} - \frac{\delta
\mu}{\hbar} + \Sigma_{11} (p=0) \right\} + \frac{(\ep)^2}{\hbar^2}\right]^{-1} \, .
\end{eqnarray}
The WT relation (\ref{WT1}) requires that this should be equal to $v$,
or equivalently
\begin{eqnarray}
\left(2gv^2+\varepsilon\bar\epsilon\right)\left(-\frac{\delta\mu}{\hbar}
+\Sigma_{11}(p=0)-\Sigma_{12}(p=0)\right)=0
\end{eqnarray}
Since the nonvanishing order parameter $v\neq0$ in the limit
$\varepsilon\rightarrow0$ is under consideration, we finally attain the
HP theorem
\begin{eqnarray}
\frac{\delta\mu}{\hbar}=\Sigma_{11}(p=0)-\Sigma_{12}(p=0) \, .
 \label{HP}
\end{eqnarray}

We again stress that the HP theorem is proven from the WT relation
in the $\varepsilon \rightarrow 0$ limit with finite $V$ here, not in
the thermodynamic (infinite volume) limit.

\subsection{The Hugenholtz-Pines theorem at one-Loop level \label{HPTOneLoop}}

In this subsection, let us check the HP theorem at one-loop level
explicitly. According to the previous subsection, the HP theorem is
a result of the WT relation. The loop expansion keeps the WT relation
at each loop level. So, we naturally expect the HP theorem
at one-loop level.

From the condition (\ref{expvphi0}) at one-loop level, $\delta \mu$
is given by
\begin{eqnarray}
\delta \mu  =  2 {\rm i} g G_{0,11} (x-x) + {\rm i} g G_{0,12} (x-x)
 =  \frac{g}{V} \sum_{{\bm p}=-\infty}^\infty  \frac{2 \epsilon_{\bm
 p} + g v^2 + 2\ep }{2 E_{\bm p}} \, .
\label{delmuexp}
\end{eqnarray}
The elements of the matrix self-energy at one-loop level are
obtained as
\begin{eqnarray}
\Sigma_{11} (p) &=& \Sigma_{22} (-p) \nn \\
& = & \frac{g^2 v^2}{\hbar^2V}
\sum^\infty_{{\bm p'}=-\infty}  \frac{1}{2 \omega_{{\bm p'}}
\omega_{{\bm p}-{\bm p'}}}
\left[ \frac{f (\omega_{\bm p'},\omega_{{\bm p}-{\bm p'}})}{\omega -
 \omega_{\bm p'} - \omega_{{\bm p}-{\bm p'}} + {\rm i} \delta} - \frac{f (-
 \omega_{\bm p'}, - \omega_{{\bm p}-{\bm p'}})}{\omega + \omega_{\bm
 p'} + \omega_{{\bm p}-{\bm p'}} - {\rm i} \delta} \right] \nn \\
& & {} + \frac{g}{\hbar^2V} \sum^\infty_{\bm p'=-\infty}
\frac{\epsilon_{\bm p'} + g v^2 +\ep}{\omega_{\bm p'}} \,
, \\
\Sigma_{12} (p) &=& \Sigma_{21} (p) \nn \\
& = & \frac{g^2 v^2}{\hbar^2V}
\sum^\infty_{{\bm p'}=-\infty} \frac{1}{2 \omega_{{\bm p'}}
\omega_{{\bm p}-{\bm p'}}}
\left[ \frac{h (\omega_{\bm p'} , \omega_{{\bm p}-{\bm p'}} )}{\omega
 - \omega_{\bm p'} - \omega_{{\bm p}-{\bm p'}} + {\rm i} \delta} - \frac{h(-
 \omega_{\bm p'} , - \omega_{{\bm p}-{\bm p'}})}{\omega + \omega_{\bm
 p'} + \omega_{{\bm p}-{\bm p'}} - {\rm i} \delta} \right] \nn \\
& &{} - \frac{g}{\hbar^2V} \sum^\infty_{{\bm p'}=-\infty}
 \frac{gv^2}{2 \omega_{\bm p'}} \, ,
\end{eqnarray}
where we have introduced functions $f (\omega_{\bm p'} , \omega_{{\bm
p}-{\bm p'}} )$ and $h (\omega_{\bm p'} , \omega_{{\bm p}-{\bm p'}}
)$ as
\begin{eqnarray}
f (\omega_{\bm p'} , \omega_{{\bm p}-{\bm p'}} ) & = & \frac{3
 \epsilon_{\bm p'} \epsilon_{{\bm p}-{\bm p'}}}{\hbar^{2}} -
 \omega_{\bm p'} \omega_{{\bm p}-{\bm p'}} + \frac{gv^2}{\hbar^2}
 (\epsilon_{\bm p'} + \epsilon_{{\bm p}-{\bm p'}}) + \frac{g^2
 v^4}{\hbar^2} \nn \\  
& &{} - \frac{g v^2}{\hbar} (\omega_{\bm p'} + \omega_{{\bm p}-{\bm
 p'}}) + \frac{1}{\hbar}(\epsilon_{\bm p'} \omega_{{\bm p}-{\bm p'}} +
 \epsilon_{{\bm p}-{\bm p'}} \omega_{\bm p'}) \nn \\
& & {} + \frac{\ep}{\hbar} \left( \frac{3 \epsilon_{\bm p'}}{\hbar} +
\frac{3 \epsilon_{{\bm p}-{\bm p'}}}{\hbar} + \frac{2 g v^2}{\hbar}
 + \omega_{\bm p'} + \omega_{{\bm p}-{\bm
p'}}\right) + \frac{3\left(\ep\right)^2}{\hbar^2} , \nn \\
\ \\ 
h (\omega_{\bm p'} , \omega_{{\bm p}-{\bm p'}}) & = & \frac{2
 \epsilon_{\bm p'} \epsilon_{{\bm p}-{\bm p'}}}{\hbar^{2}} - 2
 \omega_{\bm p'} \omega_{{\bm p}-{\bm p'}} + \frac{g^2 v^4}{\hbar^2} \nn \\
& &+ \frac{2 \ep}{\hbar} \left( \frac{\epsilon_{\bm p'}}{\hbar}
+ \frac{\epsilon_{{\bm p}-{\bm p'}}}{\hbar} \right)
+\frac{2\left(\ep\right)^2}{\hbar^2}  \, . 
\end{eqnarray}
Then, we have:
\begin{eqnarray}
& & \Sigma_{11} (p=0) \nn \\ 
& = & \frac{g^2 v^2}{\hbar^2V} \sum^\infty_{{\bm p'} = -\infty} \left(- 
\frac{1}{4 \omega_{\bm p'}^{3}} \right) \left\{ \frac{6
\epsilon_{\bm p'}^2}{\hbar^2} - 2 \omega_{\bm p'}^{2} + \frac{4 g v^2
\epsilon_{\bm p'}}{\hbar^2} + \frac{2 g^2 v^4}{\hbar^2} 
+ 4\frac{\ep}{\hbar}\left(\frac{3\epsilon_{\bm p'}}{\hbar} +
 \frac{gv^2}{\hbar} \right) 
+ \frac{6\left(\ep\right)^2}{\hbar^2} \right\} \nn \\
& & + \frac{g}{\hbar^2V}
\sum^\infty_{{\bm p'} = -\infty}  \frac{\epsilon_{\bm p'} + g v^2 +\ep }{\omega_{\bm p'}} \, , \\ 
& & \Sigma_{12} (p=0) \nn \\  
& = & \frac{g^2 v^2}{\hbar^2V} \sum^\infty_{{\bm p'} = -\infty} \left( - 
\frac{1}{4 \omega_{\bm p'}^{3}} \right) \left\{ \frac{4 \epsilon_{\bm
p'}^2}{\hbar^2} - 4 \omega_{\bm p'}^{2} + \frac{2 g^2 v^4}{\hbar^2}
+ \frac{8(\ep)\epsilon_{\bm p'}}{\hbar^2} + \frac{4\left(\ep\right)^2}{\hbar^2} \right\} \nn \\
& &- \frac{g}{\hbar^2V}\sum^\infty_{{\bm p'} = -\infty}  \frac{g
v^2}{2 \omega_{\bm p'}} \, . 
\end{eqnarray}
The following quantity, appearing in the RHS of
Eq. (\ref{HP}), is simplified as
\begin{eqnarray}
& &\Sigma_{11} (p=0) - \Sigma_{12} (p=0) \nn \\
& = & \frac{g^2 v^4}{\hbar^2V} \sum^\infty_{{\bm p'}=-\infty}  \left( -
\frac{1}{4 \omega_{\bm p'}^{3}} \right) \left( \frac{2 \epsilon_{\bm
p'}^2}{\hbar^2} + 2 \omega_{\bm p'}^{2} + \frac{4 g v^2 \epsilon_{\bm
p'}}{\hbar^2} + \frac{4\ep \epsilon_{\bm p'}}{\hbar^2} + \frac{4\ep gv^2}{\hbar^2}
 + \frac{2\left(\ep\right)^2}{\hbar^2} \right) \nn \\ 
& & {} + \frac{g}{\hbar^2V} \sum^\infty_{{\bm p'} = -\infty}  \frac{2
\epsilon_{\bm p'} + 3 g v^2 + 2\ep }{2 \omega_{\bm p'}} \nn \\
& = & - \frac{g^2 v^2}{\hbar^2V} \sum^\infty_{{\bm p'} = -\infty} 
\frac{1}{\omega_{\bm p'}} + \frac{g}{\hbar^2V} \sum^\infty_{{\bm p'} =
-\infty} \frac{2 \epsilon_{\bm p'} + 3 g v^2 +2\ep}{2
\omega_{\bm p'}} \nn \\ 
& = & \frac{g}{\hbar V} \sum^\infty_{{\bm p'} = -\infty} \frac{2
 \epsilon_{\bm p'} + g v^2 +2\ep }{2 E_{\bm p'}} \nn \\
& = & \frac{\delta \mu}{\hbar} \, .
\label{HPoneloop}
\end{eqnarray}
Thus, the HP theorem has been confirmed at one-loop level.

We comment on a possible choice of the unperturbed Hamiltonian
$H'_0$ which is obtained from replacing $\mu_0$ in Eq. (\ref{H0vphi2})
with $\mu$ given in Eq. (\ref{mudelmu}).  Then instead of $H_{\rm{int}}$
we have the perturbative Hamiltonian $H'_{\rm{int}}$ without the $\delta \mu$
term.  One can repeat the same procedure of
diagonalizing the unperturbed Hamiltonian as in Subsection \ref{DH0}.
As a result, the unperturbed Hamiltonian is given by 
\begin{eqnarray}
H_0 = \sum^\infty_{{\bm p}=-\infty} E'_{\bm p} b^{' \dag}_{\bm p}
 b'_{\bm p} + {\rm const.} \, , 
\end{eqnarray}
where
\begin{eqnarray}
E'_{\bm p}=\sqrt{\left(\epsilon_{\bm p} + g v^2 + \ep - \delta \mu
 \right)^2 - g^2  v^4} \, . \label{Eprim} 
\end{eqnarray}
We find a difficulty because $E'_{\bm p}$ becomes complex in soft momentum region
at one-loop level.  Recalling $\epsilon_{\bm p}=p^2/2m$ and taking the limit of
$\varepsilon\rightarrow0$, we have at ${\bm p}={\bm 0}$, 
\begin{eqnarray}
E'_{{\bm p}={\bm 0}} =\sqrt{\delta\mu(\delta\mu-2gv^2)} \, .
\end{eqnarray}
As is seen from Eq.~(\ref{HPoneloop}), $\delta \mu$ is determined to be positive
for positive $g$, which implies that $E'_{{\bm p}={\bm 0}}$ is pure imaginary. Thus
the choice of the unperturbed Hamiltonian $H'_0$ does not offer a consistent
treatment.

\section{Finite temperature case}

We extend the discussions on the WT relations and the HP theorem at zero
temperature to finite temperature case.  We employ the TFD formalism to describe
equilibrium situations.  As will be seen, the TFD formalism is suitable for our
purpose, because it is formulated as a canonical operator formalism of
quantum field.

\subsection{Thermofield dynamics}

In this subsection, we give a brief review of TFD formalism.
In TFD, thermal degrees of freedom are introduced by doubling each
degree of freedom through the tilde conjugation.
Thus, with an operator $A$ we associate its tilde conjugate $\tilde{A}$
according to the tilde conjugation rules \cite{U,TFD}:
\begin{eqnarray}
 (AB)\verb1~1 & = & \tilde{A} \tilde{B} \, , \\
 (c_1 A + c_2 B)\verb1~1 & = & c_1^\ast \tilde{A} + c_2^\ast \tilde{B}
  \, , \\
 (A^\dag)\verb1~1 & = & \tilde{A}^\dag \, , \\
 (\tilde{A})\verb1~1 & = & \sigma A \, , \\
 \ket{\Omega}_\beta\verb1~1 & = & \ket{\Omega}_\beta \, , \\
 {}_\beta\bra{\Omega}\verb1~1 & = & \bra{\Omega}_\beta \, ,
\end{eqnarray}
where $c_1$ and $c_2$ are complex {\it c}-numbers, $\ket{\Omega}_\beta$
and ${}_\beta\bra{\Omega}$ are the thermal vacua, and $\sigma = 1$ for
bosonic $A$, while $\sigma = - 1$ for fermionic $A$.
The total Hamiltonian $\hat{H}$
is given in terms of the non-tilde and tilde Hamiltonians as
\begin{equation}
 \hat{H} = H - \tilde{H} \, ,
\end{equation}
and the total Lagrangian $\hat{L}$ is also given as the non-tilde
Lagrangian minus the tilde one,
\begin{equation}
 \hat{L} = L - \tilde{L} \, .
\end{equation}
The thermal average is expressed as the thermal vacuum expectation value in
TFD: ${}_\beta\bra{\Omega} A \ket{\Omega}_\beta$.
For simplicity, consider one bosonic degree of freedom whose creation
and annihilation operators are represented by $a$ and $a^\dag$,
respectively. When the density matrix $\rho$ is expressed in the form of
$\rho = f^{a^\dag a}$, we have
\begin{equation}
 {}_\beta\bra{\Omega} A \ket{\Omega}_\beta = \frac{{\rm tr} [\rho A
  ]}{{\rm tr} [\rho]}
\end{equation}
and
\begin{equation}
 n = {}_\beta\bra{\Omega} a^\dag a \ket{\Omega}_\beta = \frac{f}{1-f} \,
  ,
\end{equation}
using the properties of thermal vacua.
If $f= \exp (- \beta E)$ where $\beta$ is the inverse temperature and
$E$ is an energy, ${}_\beta\bra{\Omega} A \ket{\Omega}_\beta$ represents
the thermal expectation value at finite temperature.
One can easily generalize this argument to multi-mode cases.

For convenience for following calculation, let us introduce the matrix
notation $\phi^\mu_i$ and $\bar\phi^\mu_i$ which represent tilde and
non-tilde quantum field $\varphi$ and $\tilde{\varphi}$ as
\begin{eqnarray}
\phi^1_i = \phi_i, && \phi^2_i = \tilde\phi^\dag_i \, , \nn \\
\bar\phi^1_i = \phi^\dag_i, && \bar\phi^2_i = - \tilde\phi_i \, ,
\end{eqnarray}
where
\begin{eqnarray}
\phi_1 = \varphi, && \phi_2 = \varphi^\dag \, , \nn \\
\phi_1^\dag = \varphi^\dag, && \phi_2^\dag = \varphi \, , \nn \\
\tilde\phi_1 = \varphi, && \tilde\phi_2 = \tilde\varphi^\dag \, , \nn \\
\tilde\phi_1^\dag = \tilde\varphi^\dag, && \tilde\phi_2^\dag =
 \tilde\varphi \, .
\end{eqnarray}
Then, the thermal propagators with four indices and with sixteen components
are defined by
\begin{eqnarray}
G^{\mu\nu}_{\beta,ij}(x-x') =
-{\rm i} {}_\beta \bra{\Omega} {\rm T} [ \phi^\mu_i (x) \bar\phi^\nu_j (x') ]
\ket{\Omega}_\beta \, ,
\end{eqnarray}
where $\mu, \nu = 1, 2$ and $i,j=1,2$.
The Fourier transformed thermal propagators are also defined as
\begin{equation}
 G^{\mu\nu}_{\beta,ij}(p) = \int \! \frac{{\rm d}^4 x}{(2\pi \hbar)^2}
  G^{\mu\nu}_{\beta,ij}(x) e^{\frac{{\rm i}}{\hbar} ({\bm p} \cdot {\bm x} -
  \omega t)} \, .
\end{equation}
One can derive the following relations of the Fourier transformed thermal
propagators by general properties of TFD:
\begin{eqnarray}
G^{\mu\nu}_{\beta,11} (p) = G^{\mu\nu}_{\beta,22} (-p) \, , \quad
 ~G^{\mu\nu}_{\beta,12} (p) = G^{\mu\nu}_{\beta,21} (p) \, , \quad
 G^{12}_\beta (p=0) = 0 \, .
\end{eqnarray}

\subsection{The Ward-Takahashi relation at finite temperature}

In TFD, the total Lagrangian $\hat{\mathcal L}$ density is given by
\begin{equation}
\hat{\mathcal L}={\mathcal L} - \tilde{\mathcal L} \, .
\end{equation}
We consider following continuous infinitesimal transformations:
\begin{eqnarray}
 \psi (x) & \rightarrow & \psi' (x) = \psi (x) + \xi \delta \psi (x) \nn
  \\
 \tilde{\psi} (x) & \rightarrow & \tilde{\psi}' (x) = \tilde{\psi} (x) +
  \xi \delta \tilde{\psi} (x) \, . \label{transTFD}
\end{eqnarray}
The N{\" o}ther theorem at finite temperature is obtained from the
infinitesimal transformations (\ref{transTFD}) as
\begin{eqnarray}
\frac{\partial}{\partial t} \hat{N}^0 (x) + \nabla \cdot {\hat{\bm N}}
 (x) = \delta \hat{\mathcal L} \, ,
\end{eqnarray}
where
\begin{eqnarray}
\hat{N}^\mu (x) = \frac{\partial{\mathcal L}}{\partial \left(
\partial_\mu \Psi (x) \right)} \delta \Psi(x) - \frac{\partial
\tilde{\mathcal L}}{\partial \left( \partial_\mu \tilde\Psi(x) \right)}
\delta \tilde\Psi(x)
\end{eqnarray}
and
\begin{eqnarray}
\delta \hat{\mathcal L}=\delta {\mathcal L} - \delta \tilde{\mathcal L}
 \, .
\end{eqnarray}
Here, $\delta {\mathcal L}$ was defined in Eq. (\ref{delLdef}) and $\delta
\tilde{\mathcal L}$ is defined as
\begin{equation}
\xi \delta \tilde{\mathcal L} = \tilde{\mathcal L} [\tilde{\Psi}' (x)] -
 \tilde{\mathcal L} [\tilde{\Psi} (x)] \, .
\end{equation}
One can easily obtain the WT relation at finite temperature by replacing
$\delta {\mathcal L}$ with $\delta \hat{\mathcal L}$ in
Eq. (\ref{WTgeneral}):
\begin{eqnarray}
& &\sum_{a = 1}^{n} {\rm i}\hbar {}_\beta \bra{\Omega} {\rm T} [ \Psi (x_1) \cdots
 \delta \Psi (x_a) \cdots \Psi (x_n) ] \ket{\Omega}_\beta \nn \\
&=& \int \! {\rm d}^4
 x {}_\beta \bra{\Omega} {\rm T} [ \delta \hat{\mathcal L} (x) \Psi (x_1) \cdots
 \Psi (x_n) ] \ket{\Omega}_\beta  \, . \label{WTtemp}
\end{eqnarray}

Now we consider the following WT relation with respect to the global
phase transformation, i.e., $\delta \Psi (x) = {\rm i} \Psi (x)$ and
$\delta \tilde{\Psi} (x) = - {\rm i} \tilde{\Psi} (x)$ in
Eq. (\ref{transTFD}).
To deal with the spontaneous breakdown of the symmetry at finite
temperature, we also introduce an infinitesimal
symmetry breaking term with respect to the tilde field as
\begin{equation}
 \tilde{\mathcal L}_\varepsilon = (\ep) v \left[ \tilde{\Psi} (x) +
 \tilde{\Psi}^\dag (x) \right] \, ,
\end{equation}
which is a tilde conjugate of ${\mathcal L}_\varepsilon$ (\ref{isbt}).
Our starting total Lagrangian density at finite temperature is given by
\begin{equation}
\hat{\mathcal L}_{\rm tot} = {\mathcal L}_{\rm tot} - \tilde{\mathcal
 L}_{\rm tot} \, ,  \label{totLTFD}
\end{equation}
where ${\mathcal L}_{\rm tot}$ was defined in Eq. (\ref{Ltotdef}) and
$\tilde{\mathcal L}_{\rm tot}$ is defined as
\begin{equation}
 \tilde{\mathcal L}_{\rm tot} = \tilde{\mathcal L} + \tilde{\mathcal
 L}_\varepsilon \, .
\end{equation}

The N{\" o}ther charge of the total Lagrangian $\hat{\mathcal L}_{\rm tot}$
(\ref{totLTFD}) with respect to the global phase transformations is
obtained:
\begin{eqnarray}
\hat{N} (t) & = & N(t) - \tilde{N} (t) = - \int \! d^3x \left\{ \Psi^\dag (x) \Psi (x) + \tilde\Psi^\dag
(x) \tilde\Psi (x) \right\} \, .
\end{eqnarray}
This generates the infinitesimal transformation, e.g.,
\begin{eqnarray}
\delta {\mathcal L}_{\rm tot} & = & {\rm i}[\hat{N} (t) , {\mathcal L}_{\rm tot}
 (x) ]  = {\rm i} (\ep) v \left[ \Psi (x) - \Psi^\dag (x) \right] \, ,
 \label{delLtot} \\
\delta \delta {\mathcal L}_{\rm tot} & = &{\rm i} [ \hat{N} (t), \delta {\mathcal
 L}_{\rm tot} (x) ] =  - (\ep) v \left[ \Psi (x) + \Psi^\dag (x) \right]
 \, , \label{deldelLtot}
\end{eqnarray}
and
\begin{equation}
\delta \hat{\mathcal L}_{\rm tot} (x) =  {\rm i}[\hat{N}(t), \hat{\mathcal
 L}_{\varepsilon} (x)] = {\rm i} (\ep) v \left[ \Psi (x) -  \Psi^\dag (x)
 +\tilde\Psi(x) -\tilde\Psi^\dag(x) \right] \, . \label{delbarLtot}
\end{equation}
One can derive the following restricted relation from the WT relation at
finite temperature (\ref{WTtemp}),  and from
 Eqs. (\ref{delLtot}), (\ref{deldelLtot}) and (\ref{delbarLtot}):
\begin{eqnarray}
i \hbar {}_\beta \bra{\Omega} \delta \delta {\mathcal L}_{\rm tot} (x)
 \ket{\Omega}_\beta  = \int \! {\rm d}^4 x'\, {}_\beta \bra{\Omega} {\rm T} [ \delta
 \hat{\mathcal L}_{\rm tot} (x') \delta {\mathcal L}_{\rm tot} (x) ]
 \ket{\Omega}_\beta \, . \label{WTdelLFT}
\end{eqnarray}
We rewrite this relation  in terms of the thermal propagators:
\begin{eqnarray}
v & = & - \frac{(\ep)v}{2 \hbar} \int d^4 x' \Big[ G^{11}_{\beta,11} (x-x')
+ G^{11}_{\beta,22} (x-x') - G^{11}_{\beta,12} (x-x') -
G^{11}_{\beta,21} (x-x') \nn \\
& & {} + G^{12}_{\beta,11} (x-x') + G^{12}_{\beta,22} (x-x') -
G^{12}_{\beta,12} (x-x') - G^{12}_{\beta,21} (x-x') \Big] \nn \\
& = & - \frac{(\ep)v}{\hbar} \left[ G^{11}_{\beta,11} (p=0) -
G^{11}_{\beta,12} (p=0) \right] \label{WTtemp2} \, .
\end{eqnarray}

\subsection{The Ward-Takahashi relation at tree level in finite
  temperature case}

Let us introduce the operators $\{\xi_{\bm p},\xi^\dag_{\bm p}\}$
and $\{\tilde\xi_{\bm p}, \tilde\xi^\dag_{\bm p}\}$ related to the
operators $\{b_{\bm p},b_{\bm p}^\dag\}$ and $\{{\tilde
b}_{\bm p},{\tilde b}^\dag_{\bm p}\}$ by the following thermal
Bogoliubov transformation:
\begin{eqnarray}
b_{\bm p} = c_{\bm p} \xi_{\bm p} + s_{\bm p}
 {\tilde\xi}^\dag_{\bm p} \, ,\label{i.t}
\end{eqnarray}
with
\begin{eqnarray}
c_{\bm p} = \frac{1}{\sqrt{\mathstrut 1 - e^{-\beta E_{\bm p}}}} \, ,
 \quad s_{\bm p} = \frac{e^{-\frac{\beta E_{\bm p}}{2}}} {\sqrt{\mathstrut
 1-e^{-\beta E_{\bm p}}}} \, ,
\end{eqnarray}
where the energy $E_{\bm p}$ is given in Eq. (\ref{Ep}).
Since $c_{\bm p}^2 - s_{\bm p}^2 = 1$, we  find that
$[\xi_{\bm p},\xi_{\bm p'}^\dag] = [\tilde \xi_{\bm p},
\tilde\xi^\dag_{\bm p'}] = \delta_{{\bm p}{\bm p'}}$ and the other commutation
relations vanish.
With these operators, the unperturbed Hamiltonian at the finite
temperature in TFD is given as
\begin{eqnarray}
\hat{H}_0 = \sum_{{\bm p}=-\infty}^{\infty} E_{\bm p} \left( b^\dag_{\bm
 p} b_{\bm p} - {\tilde b}^\dag_{\bm p} {\tilde b}_{\bm p}\right)
 = \sum_{{\bm p}=-\infty}^{\infty} E_{\bm p} \left( \xi^\dag_{\bm p}
\xi_{\bm p} -\tilde\xi^\dag_{\bm p} \tilde\xi_{\bm p} \right) \, .
\end{eqnarray}

The field operator $\varphi(x)$ is rewritten in terms of the
operators $\{\xi_{\bm p},\xi^\dag_{\bm p}\}$ and
$\{\tilde\xi_{\bm p},\tilde\xi^\dag_{\bm p}\}$ as
\begin{eqnarray}
\varphi(x) &=& \frac{1}{\sqrt{\mathstrut V}}  \sum^{\infty}_{{\bm p}=-\infty}
 \left[ \left( c_{\bm p} \xi_{\bm p} +
 s_{\bm p} \tilde\xi^\dag_{\bm p} \right)u_{\bm p} e^{\frac{{\rm i}}{\hbar}
 ({\bm p} \cdot {\bm x} - E_{\bm p} t)} \right. \nn \\
& &{} +\left. \left( c_{\bm p} \xi^\dag_{\bm
 p} + s_{\bm p} \tilde\xi_{\bm p} \right) v_{\bm p}
 e^{-\frac{{\rm i}}{\hbar} ({\bm p} \cdot {\bm x} - E_{\bm p} t)} \right] \, .
\end{eqnarray}
We construct the unperturbed propagators with this field operator and
its Hermite and tilde conjugate:
\begin{eqnarray}
G^{\mu\nu}_{\beta,0,ij} (x - x') = - {\rm i} {}_\beta \bra{\Omega_0} {\rm T} [ \phi^\mu_i (x)
 \bar\phi^\nu_j (x') ] \ket{\Omega_0}_\beta \, ,
\end{eqnarray}
and its Fourier transformed one
\begin{eqnarray}
G^{\mu\nu}_{\beta,0,ij} (p) = \int \frac{{\rm d}^4x}{\left( 2 \pi \hbar
\right)^2} G^{\mu\nu}_{\beta,0,ij} (x) e^{\frac{{\rm i}}{\hbar} ({\bm p}
\cdot {\bm x} - \omega t)} \, .
\end{eqnarray}
Here, the unperturbed thermal vacuum $\ket{\Omega_0}_\beta$ is specified by
$\xi_{\bm p}\ket{\Omega_0}_\beta = \tilde\xi_{\bm p} \ket{\Omega_0}_\beta = 0$.
The explicit forms of the matrix elements, which are necessary to check
 the WT relation, are given as
\begin{eqnarray}
G^{11}_{\beta,0,11} (p) & = & G^{11}_{\beta,0,22} (-p) \nn \\
& = & \left(\frac{c^2_{\bm p}}{\omega - \omega_{\bm p} + {\rm i} \delta} -
\frac{s^2_{\bm p}}{\omega - \omega_{\bm p} - {\rm i} \delta} \right) u^2_{\bm
p} \nn \\
& &+ \left(\frac{s^2_{\bm p}}{\omega + \omega_{\bm p} + {\rm i} \delta} -
\frac{c^2_{\bm p}}{\omega + \omega_{\bm p} - {\rm i} \delta} \right) v^2_{\bm
p} \label{Gbeta011} \\
G^{11}_{\beta,0,12} (p) & = & G^{11}_{\beta,0,21} (p) \nn \\
& = & \left(\frac{c^2_{\bm p}}{\omega - \omega_{\bm p} + {\rm i} \delta} +
\frac{s^2_{\bm p}}{\omega + \omega_{\bm p} + {\rm i} \delta} \right. \nn \\
& & - \left. \frac{c^2_{\bm p}}{\omega + \omega_{\bm p} - {\rm i} \delta} - \frac{s^2_{\bm p}}{\omega -
\omega_{\bm p} - {\rm i} \delta} \right) u_{\bm p} v_{\bm p} \, .
\label{Gbeta012}
\end{eqnarray}
At $p = 0$ they are reduced to
\begin{eqnarray}
G^{11}_{\beta,0,11} ( p = 0 ) & = & G^{11}_{\beta,0,22} ( p = 0 ) \nn \\
& = & - \frac{1}{\omega_{{\bm p} = {\bm 0}}} \left(c_{{\bm p} = {\bm
0}}^2 - s^2_{{\bm p} = {\bm 0}} \right) \left(u_{{\bm p} = {\bm 0}}^2 +
v^2_{{\bm p} = {\bm 0}} \right) \nn \\
& = & - \frac{u_{{\bm p} = {\bm 0}}^2 + v_{{\bm p} = {\bm
0}}^2}{\omega_{{\bm p}={\bm 0}}} 
= G_{0,11} ( p = 0 )  =  G_{0,22} ( p = 0 )\label{Gbeta0p011} \, , \\
G^{11}_{\beta,0,12} ( p = 0 ) & = & G^{11}_{\beta,0,21} ( p = 0 ) \nn \\
& = & -\frac{2}{\omega_{{\bm p} = {\bm 0}}} \left( c_{{\bm p} = {\bm
0}}^2 - s_{{\bm p} = {\bm 0}}^2 \right) u_{{\bm p} = {\bm 0}} v_{{\bm p}
= {\bm 0}} \nn \\
& = & - \frac{2u_{{\bm p} = {\bm 0}}v_{{\bm p} = {\bm 0}}}{\omega_{{\bm
 p}={\bm 0}}} 
= G_{0,12} ( p = 0 ) = G_{0,21} ( p = 0 ) \, . \label{Gbeta0p012}
\end{eqnarray}
Substituting Eqs. (\ref{Gbeta0p011}) and (\ref{Gbeta0p012}) into 
Eq. (\ref{WTtemp2}), we find that the WT relation at finite temperature
holds at tree level:
\begin{equation}
- \frac{(\ep)v}{\hbar} \left[ G^{11}_{\beta,11} (p = 0) - G^{11}_{\beta,12}
(p = 0) \right]
= - \frac{(\ep)v}{\hbar} \left[ G_{11} ( p = 0 ) - G_{12} ( p = 0 ) \right]
= v
\end{equation}
In other words, at tree level, the Goldstone theorem also holds at
finite temperature.

\subsection{The Hugenholtz-Pines theorem at finite temperature}

In Subsection \ref{HPTZT}, the HP theorem at $T=0$ was derived from the
WT relation.
Now we will generalize this story to finite temperature case.
We considered the quantum corrections of the chemical potential (\ref{mudelmu}) at $T=0$.
At finite temperature one has to take account of thermal corrections
as well as quantum ones.
Explicitly $\delta \mu$ in Eq. (\ref{mudelmu}) now represents
 the thermal and quantum corrections to the
chemical potential at tree level.

We consider the following Schwinger--Dyson equation at finite
temperature:
\begin{eqnarray}
G_\beta^{-1}(p)=G^{-1}_{\beta,0}(p)-\Sigma_\beta(p) + \frac{\delta \mu}{\hbar} I_{4\times 4} \, ,
\end{eqnarray}
where $G_\beta(p)$ and $\Sigma_\beta(p)$, having the indices $(\mu,\nu)$
and $(i,j)$ are the full propagator and self-energy
at finite temperature, respectively, and $I_{4\times4}$ is a unity $4\times4$ matrix.
In TFD, one can find the following relations in general,
\begin{eqnarray}
G^{11}_\beta (p = 0) = G^{22}_\beta (p = 0) \, , && G^{12}_\beta (p = 0)
 = G^{21}_\beta (p = 0) = 0 \, , \nn \\
\Sigma^{11}_\beta (p = 0) = \Sigma^{22}_\beta (p = 0) \, , &&
\Sigma^{12}_\beta (p = 0) = \Sigma^{21}_\beta (p = 0) = 0 \, , \nn \\
G^{11}_{\beta,11} (p = 0) = G^{11}_{\beta,22} (p = 0) \, , &&
 G^{11}_{\beta,12} (p = 0) = G^{11}_{\beta,21} (p = 0) \, , \nn \\
\Sigma^{11}_{\beta,11} (p = 0) = \Sigma^{11}_{\beta,22} (p = 0) \, , &&
\Sigma^{11}_{\beta,12} (p = 0) = \Sigma^{11}_{\beta,21} (p = 0) \, .
\end{eqnarray}

Let us calculate the RHS of Eq. (\ref{WTtemp2}),
\begin{eqnarray}
&& - \frac{(\ep) v}{\hbar} \left( G^{11}_{\beta,11} (p = 0) -
G^{11}_{\beta,12} (p = 0) \right) \nn \\
& = & - \frac{(\ep) v}{\hbar} \left[ \left\{ {G^{-1}_\beta}^{11}_{11} (p =
0) \right\}^2 - \left\{ {G^{-1}_\beta}^{11}_{12} (p = 0) \right\}^2
\right]^{-1} \left[ {G^{-1}_\beta}^{11}_{11} (p = 0) +
{G^{-1}_\beta}^{11}_{12} (p = 0) \right] \nn \\
& = & - \frac{(\ep) v}{\hbar} \left[ \left\{ {G^{-1}_{\beta,0}}^{11}_{11}
(p = 0) - \Sigma^{11}_{\beta,11} (p = 0) +\frac{\delta \mu }{\hbar} \right\}^2 - \left\{
{G^{-1}_{\beta,0}}^{11}_{12} (p = 0) - \Sigma^{11}_{\beta,12} (p = 0)
\right\}^2 \right]^{-1} \nn \\
&& \times \left[ {G^{-1}_{\beta,0}}^{11}_{11} (p = 0) + {G^{-1}_{\beta, 0}}^{11}_{12} (p = 0) - \Sigma^{11}_{\beta,11} (p = 0) -
\Sigma^{11}_{\beta,12} (p = 0)  +\frac{\delta \mu}{\hbar} \right] \, .
\end{eqnarray}
From Eqs. (\ref{Gbeta0p011}) and (\ref{Gbeta0p012}), we find that this expression
is very similar in form to that at $T=0$ in Eq. (\ref{WTloop}),
except that the self-energy terms $\Sigma^{11}_{\beta,11}(p=0)$ and
$\Sigma^{11}_{\beta,12}(p=0)$ replace $\Sigma_{11}(p=0)$ and
$\Sigma_{12}(p=0)$ there.
Thus the WT relation at finite temperature is equivalent to
the following HP theorem at finite temperature:
\begin{eqnarray}
\frac{\delta \mu}{\hbar} = \Sigma^{11}_{\beta,11} (p = 0) -
\Sigma^{11}_{\beta,12} (p = 0) \, , \label{HPtemp}
\end{eqnarray}
which is very similarly as Eq. (\ref{HP}) has been derived at zero-temperature.

\subsection{The Hugenholtz-Pines theorem at one-loop level in finite
  temperature case}

Finally, in an explicit calculation at one-loop level
in finite temperature case,
we check the conclusion in the previous subsection, i.e.,
the HP theorem or the WT relation.
The condition (\ref{expvphi0}) fixes $\delta\mu$ at one-loop level in
finite temperature case (See Eq. (\ref{delmuexp}) at $T=0$),
\begin{eqnarray}
\delta \mu & = & 2 {\rm i} g G_{\beta,0,11} (x-x) + {\rm i} g G_{\beta,0,12}
 (x-x) \nn\\
& = & g \sum^\infty_{{\bm p'}=- \infty} \frac{1}{V} \coth \left(
\frac{\beta E_ {\bm p'}}{2} \right) \frac{2 \epsilon_{\bm p'} + g v^2 +
2 \ep }{2 E_{\bm p'}} \, .
\end{eqnarray}
To investigate the HP theorem, it is necessary to have
only the two
matrix elements of the self-energy, $\Sigma^{11}_{\beta,11}(p)$ and
$\Sigma^{11}_{\beta,12}(p)$. They read as
\begin{eqnarray}
\Sigma^{11}_{\beta,11}(p)
& = & \frac{g^2 v^2}{\hbar^2}\sum_{{\bm p}=-\infty}^\infty \frac{1}{V}
\frac{1}{2\omega_{\bm p'} \omega_{\bm{p-p'}}} \nn \\
& & \times \Bigg[ \left(c^{2}_{\bm p'} c^{2}_{\bm {p-p'}}
 - s^{2}_{\bm p'} s^{2}_{\bm {p-p'}} \right)
\left\{ \frac{f (\omega_{\bm p'},\omega_{{\bm p}-{\bm p'}})}{\omega -
 \omega_{\bm p'} - \omega_{{\bm p}-{\bm p'}} + {\rm i} \delta} - \frac{f (-
 \omega_{\bm p'}, - \omega_{{\bm p}-{\bm p'}})}{\omega + \omega_{\bm
 p'} + \omega_{{\bm p}-{\bm p'}} - {\rm i} \delta} \right\} \nn \\
& & + \left(c^{2}_{\bm p'} s^{2}_{\bm {p-p'}} - s^{2}_{\bm p'} c^{2}_{\bm {p-p'}} \right)
\left\{ \frac{f (\omega_{\bm p'},\omega_{{\bm p}-{\bm p'}})}{\omega -
 \omega_{\bm p'} + \omega_{{\bm p}-{\bm p'}} + {\rm i} \delta} - \frac{f (-
 \omega_{\bm p'}, - \omega_{{\bm p}-{\bm p'}})}{\omega + \omega_{\bm
 p'} - \omega_{{\bm p}-{\bm p'}} - {\rm i} \delta} \right\} \Bigg] \nn \\
& & + \frac{g}{\hbar^2} \sum^\infty_{{\bm p'}=- \infty} \frac{1}{V}
 \coth \left( \frac{\beta E_ {\bm p'}}{2} \right)
\frac{\epsilon_{\bm p'} + g v^2 + \ep }{\omega_{\bm p'}},
\end{eqnarray}
\begin{eqnarray}
\Sigma^{11}_{\beta,12}(p)
& = & \frac{g^2 v^2}{\hbar^2}\sum^\infty_{{\bm p'}=- \infty} \frac{1}{V}
\frac{1}{2\omega_{\bm p'} \omega_{\bm{p-p'}}} \nn \\
& & \times \Bigg[ \left(c^{2}_{\bm p'} c^{2}_{\bm {p-p'}}
- s^{2}_{\bm p'} s^{2}_{\bm {p-p'}} \right)
\left\{ \frac{h (\omega_{\bm p'},\omega_{{\bm p}-{\bm p'}})}{\omega -
 \omega_{\bm p'} - \omega_{{\bm p}-{\bm p'}} + {\rm i} \delta} - \frac{h (-
 \omega_{\bm p'}, - \omega_{{\bm p}-{\bm p'}})}{\omega + \omega_{\bm
 p'} + \omega_{{\bm p}-{\bm p'}} - {\rm i} \delta} \right\} \nn \\
& & {} + \left(c^{2}_{\bm p'} s^{2}_{\bm {p-p'}} -
 s^{2}_{\bm p'} c^{2}_{\bm {p-p'}} \right)
\left\{ - \frac{h (\omega_{\bm p'},\omega_{{\bm p}-{\bm p'}})}{\omega -
 \omega_{\bm p'} + \omega_{{\bm p}-{\bm p'}} + {\rm i} \delta} + \frac{h (-
 \omega_{\bm p'}, - \omega_{{\bm p}-{\bm p'}})}{\omega + \omega_{\bm
 p'} - \omega_{{\bm p}-{\bm p'}} - {\rm i} \delta} \right\} \Bigg] \nn \\
& &{} - \frac{g}{\hbar^2} \sum^\infty_{{\bm p'}=- \infty} \frac{1}{V}
 \coth \left( \frac{\beta E_ {\bm p'}}{2} \right) \frac{gv^2}{2
 \omega_{\bm p'}} \, ,
\end{eqnarray}
Using $c^{2}_{\bm p}-s^{2}_{\bm p}=1$ and referring to
the calculations in Subsection \ref{HPTOneLoop},
we can easily find that the RHS of Eq. (\ref{HPtemp}) is
given by
\begin{eqnarray}
\Sigma^{11}_{\beta,11} (p=0) - \Sigma^{11}_{\beta,12} (p=0) & = &
\frac{g}{\hbar} \sum^\infty_{{\bm p'} = - \infty} \frac{1}{V} \coth
\left( \frac{\beta E_ {\bm p'}}{2} \right) \frac{2 \epsilon_{\bm p'} +
g v^2 + 2\ep }{2 E_{\bm p'}} \nn \\
& = & \frac{\delta \mu}{\hbar} \, .
\end{eqnarray}
Thus it has been confirmed that the HP theorem or the WT relation
is satisfied at one-loop level. In other words, our approximate calculation
scheme respects the Goldstone theorem at the one-loop level
in finite temperature case.

\section{Summary}

We constructed an approximate scheme satisfying the Goldstone theorem for
Bose--Einstein condensed gas in a box with the periodic boundary
condition and generalized it to finite temperature case using TFD.
When energy spectrum is continuous as in homogeneous situation,
one takes the thermodynamic limit and then
the Bogoliubov's prescription do not cause serious problem for the
Goldstone theorem, since the NG mode, which is ${\bm p} = {\bm 0}$ mode,
is just a point in momentum integral and omitting it does not affect the result.
On the other hands, when a system has a finite volume
so that energy spectrum is discrete and that the concept of the
thermodynamic limit is not applicable, we found that it is necessary to deal with the NG
mode explicitly for satisfying the
WT relation, that is, the Goldstone theorem.
Without the thermodynamic limit, the HP theorem is also derived from the
WT relation.
Furthermore the NG mode contributes essentially to the unitarily
inequivalence of vacua in the case of the SSB for finite volume system.
Then it is found that the scheme for the trapped dilute Bose system
satisfying the Goldstone theorem also need explicit treatment of the NG
mode.

\section*{Acknowledgments}
The authors would like to thank Professor I.~Ohba and Professor H.~Nakazato
for helpful comments and encouragements,
and Dr. M.~Mine for useful discussions.
The authors thank the Yukawa Institute for Theoretical Physics
at Kyoto University for offering us the opportunity of discussions
during the YITP workshop YITP-W-05-09
on ``Thermal Quantum Field Theories and Their Applications,''
useful to complete this work.
This work is partly supported by a Grant-in-Aid for Scientific Research
(C) (No.~17540364) from the Japan Society for the Promotion of Science,
for Young Scientists (B) (No.~17740258) and for
Priority Area Research (B) (No.~13135221) both from the Ministry of
Education, Culture, Sports, Science and Technology, Japan.
M.~O. and Y.~Y. wish to express their thanks for Waseda University Grant
for Special Research Projects (2005).

\appendix

\section{Bogoliubov transformation and vacua\label{AppA}}

In this appendix, we give explicit relation between $\KV$ and
$\ket{\Omega_0}$ which are the vacuum of $a_{\bm p}$ and $b_{\bm p}$
respectively \cite{Finite}.  

The Bogoliubov transformation (\ref{BTtree}) is written as the following
matrix form: 
\begin{equation}
 \left(
  \begin{array}{c}
   {b}_{\bm p} \\
   {b}_{- {\bm p}}^\dag
  \end{array}
 \right) 
 = 
 \left(
  \begin{array}{cc}
   u_{\bm p} & - v_{\bm p} \\
   - v_{\bm p} & u_{\bm p}
  \end{array}
 \right)
 \left(
  \begin{array}{c}
   {a}_{\bm p} \\
   {a}_{-{\bm p}}^\dag
  \end{array}
 \right) 
 \, . \label{yXYx}
\end{equation}
Next, we introduce an operator $U_{\bm p}$ which induces the
transformation (\ref{yXYx}): 
\begin{equation}
 {b}_{\bm p} = U_{\bm p} {a}_{\bm p} {U}_{\bm p}^\dag \, , \quad
 {b}_{-{\bm p}}^\dag =  U_{\bm p} {a}_{-{\bm p}}^\dag U_{\bm p}^\dag \, ,
  \label{ySxS} 
\end{equation}
where $U_{\bm p}$ must be unitary,
\begin{equation}
 U_{\bm p} U_{\bm p}^\dag = U_{\bm p}^\dag U_{\bm p} = \hat{1} \, .
  \label{SS1} 
\end{equation}
Here, $\hat{1}$ is the identity operator.
Eqs. (\ref{ySxS}) imply 
\begin{equation}
 b_{\bm p} U_{\bm p} \KV = 0 \, , \quad \BV U_{\bm p}^\dag
 b_{-{\bm p}}^\dag = 0 \, .  \label{ySvac} 
\end{equation}
We define the unitary operator $U$ as 
\begin{equation}
 U = \prod_{{\bm p} = {\bm 0}}^\infty U_{\bm p} \, , 
\end{equation}
where $\prod_{{\bm p} = {\bm 0}}^\infty$ contains the suffix ${\bm
p} = {\bm 0}$, and $U$ is obviously unitary. 
Then the relation between $\ket{\Omega_0}$ and $\KV$ is given as 
\begin{equation}
 \ket{\Omega_0} = U \KV \, .  \label{OmegaU0}
\end{equation}

First, we consider the ${\bm p} \ne {\bm 0}$ part of $U$. 
To determine an explicit representation of $U_{\bm p}$ in terms of
$a_{\bm p}$ and ${a}_{-{\bm p}}^\dag$, we define the coherent state of
$a_{\bm p}$ and $a_{-{\bm p}}$, 
\begin{equation}
\ket{z({\bm p})} = \exp [ z_{\bm p} {a}_{\bm p}^\dag] \exp [ z_{-{\bm p}}
 {a}_{-{\bm p}}^\dag] \ket{0} = 
 \left[ \sum_{\ell=0}^\infty \frac{\left( z_{\bm p} {a}_{\bm p}^\dag
\right)^\ell}{\ell !} \right] 
\left[ \sum_{\ell'=0}^\infty \frac{\left( z_{-{\bm p}} {a}_{-{\bm
 p}}^\dag \right)^{\ell'}}{\ell' !} \right]
\ket{0} \, , \label{defcoh}
\end{equation}
where $z_{\bm p}$ and $z_{-{\bm p}}$ are eigenvalues of ${a}_{\bm p}$
and $a_{-{\bm p}}$, i.e., 
\begin{equation}
 {a}_{\bm p} \ket{z({\bm p})} = z_{\bm p} \ket{z({\bm p})} \, , \quad  
 a_{-{\bm p}} \ket{z({\bm p})} = z_{-{\bm p}} \ket{z({\bm p})} \, .
 \label{xz} 
\end{equation}
The coherent state has the properties of  
\begin{eqnarray}
 {a}_{\bm p}^\dag \ket{z({\bm p})} & = & \partial_{z_{\bm p}}
  \ket{z({\bm p})} \, , \quad  
 {a}_{-{\bm p}}^\dag \ket{z({\bm p})} = \partial_{z_{-{\bm p}}}
 \ket{z({\bm p})} \, , \\ 
 \bra{z({\bm p})} {a}_{\bm p}^\dag & = & \bra{z({\bm p})} z_{\bm p}^\ast
  \, , \quad  
 \bra{z({\bm p})} {a}_{-{\bm p}}^\dag = \bra{z({\bm p})} z_{-{\bm
 p}}^\ast \, ,  \\ 
 \bra{z({\bm p})} {a}_{\bm p} & = & \partial_{z_{\bm p}^\ast}\bra{z({\bm
  p})}  \, , \quad  
 \bra{z({\bm p})} {a}_{-{\bm p}} = \partial_{z_{-{\bm
 p}}^\ast}\bra{z({\bm p})}  \, ,  
\end{eqnarray}
and 
\begin{equation}
 \braket{0}{z({\bm p})} = 1 \, .
\end{equation}
The completeness condition of the coherent state reads 
\begin{equation}
\int \! {\rm d} \mu (z({\bm p})) \, \ket{z({\bm p})} \bra{z({\bm p})} = \int
 \! {\rm d} \mu (z({\bm p})) \, \left[ \ket{z_{\bm p}} \bra{z_{\bm p}} \otimes
 \ket{z_{-{\bm p}}} \bra{z_{-{\bm p}}} \right] = \hat{1} \, , \label{cmpz} 
\end{equation}
where the measure is defined as 
\begin{eqnarray}
{\rm d} \mu (z({\bm p})) & = &
\left[ e^{- |z_{\bm p}|^2} \frac{{\rm d} x_{\bm p} \, {\rm d} y_{\bm p}}{ \pi }
\right] 
\left[ e^{- |z_{-{\bm p}}|^2} \frac{{\rm d} x_{-{\bm p}} \, {\rm d} y_{-{\bm p}}}{
 \pi } \right] \, , \\
 z_{\pm {\bm p}} & = & x_{\pm {\bm p}}+{\rm i} y_{\pm {\bm p}} \, .  
\end{eqnarray}

We also mention the following formula of 
a generalized Gaussian integral: 
\begin{eqnarray}
&& \int \! {\rm d} \mu (z({\bm p})) \, \exp \left[ - \frac{1}{2} \left( z^{\rm
 T} ({\bm p}) M z ({\bm p}) + z^{\ast {\rm T}} ({\bm p}) N^\ast z^\ast
 ({\bm p}) \right) + u^{\rm T} z({\bm p}) + v^{\ast {\rm T}} z^\ast
({\bm p}) \right] \nn \\ 
& = & \left[ \det (I - M N^\ast) \right]^{-\frac{1}{2}} \nn \\
 && \times \exp \biggl[ - \frac{1}{2} u^{\rm T} ( I - N^\ast M)^{-1}
 N^\ast u - \frac{1}{2} v^{\ast {\rm T}} (I - MN^\ast)^{-1} M v^\ast
 \nn \\
&& {}\qquad \qquad + v^{\ast {\rm T}} (I - MN^\ast)^{-1} u \biggr] \, .
 \label{genGaussInt} 
\end{eqnarray}
Here, $M$ and $N^\ast$ are $2 \times 2$ symmetric matrices, 
\begin{equation}
 M^{\rm T} = M \, , \quad N^{\ast {\rm T}} = N^\ast \, , 
\end{equation}
$I$ is a unity $2\times2$ matrix, and $z$, $u$ and $v$ are {\it c}-number 2-components vectors, 
\begin{equation}
 z ({\bm p}) = 
 \left(
  \begin{array}{c}
   z_{\bm p} \\
   z_{-{\bm p}} 
  \end{array}
 \right) \, , \qquad
 u = 
 \left(
  \begin{array}{c}
   u_+ \\
   u_- 
  \end{array}
 \right) \, , \qquad
 v = 
 \left(
  \begin{array}{c}
   v_+ \\
   v_- 
  \end{array}
 \right) \, . 
\end{equation}

Now, we are led to the operator $U_{\bm p}$ explicitly in terms
of ${a}_{\bm p}$ and ${a}^\dag_{-{\bm p}}$.  One can easily derive the
following relations from Eqs. (\ref{ySxS})--(\ref{ySvac})
and (\ref{xz}): 
\begin{equation}
 ( {b}_{\bm p} - z_{\bm p} ) {U}_{\bm p} \ket{z({\bm p})} = 0 \, , \quad  
 ( {b}_{-{\bm p}}^\dag - \partial_{z_{-{\bm p}}} ) {U}_{\bm p}
 \ket{z({\bm p})} = 0 \, .  \label{yzS}
\end{equation}
Putting  the expression (\ref{yXYx}) into Eqs.~(\ref{yzS}), we obtain 
\begin{eqnarray}
 (u_{\bm p} {a}_{\bm p} - v_{\bm p} {a}_{-{\bm p}}^\dag - z_{\bm p})
  {U}_{\bm p} \ket{z({\bm p})} & = & 0  \, , \\ 
 (- v_{\bm p} {a}_{\bm p} + u_{\bm p} {a}_{-{\bm p}}^\dag -
  \partial_{z_{-{\bm p}}}) {U}_{\bm p} \ket{z({\bm p})} & = & 0 \, , 
\end{eqnarray}
whose inner product with the coherent state $\bra{w({\bm p})}$ are given
by 
\begin{eqnarray}
 (u_{\bm p} \partial_{w_{\bm p}^\ast} - v_{\bm p} w_{-{\bm p}}^\ast -
  z_{\bm p}) \bra{w({\bm p})} U_{\bm p} \ket{z({\bm p})} & = & 0 \, , \\  
 (- v_{\bm p} \partial_{w_{\bm p}^\ast} + u_{\bm p} w_{-{\bm p}}^\ast - 
 \partial_{z_{-{\bm p}}}) \bra{w({\bm p})} U_{\bm p} \ket{z({\bm p})} &
 = & 0 \, . 
\end{eqnarray}
One finds that the following solution is admitted, 
\begin{eqnarray}
& & \bra{w({\bm p})} U_{\bm p} \ket{z({\bm p})} \nn \\
&=& \frac{1}{u_{\bm p}}
 \exp \left[ \frac{1}{2} w^{\ast {\rm T}} ({\bm p}) X_{\bm p} w^\ast
 ({\bm p}) - \frac{1}{2} z^{\rm T} ({\bm p}) X_{\bm p} z ({\bm p}) +
 w^{\ast {\rm T}} ({\bm p}) Y_{\bm p} z ({\bm p}) \right] \, , 
 \label{tSz} 
\end{eqnarray}
where
\begin{equation}
 X_{\bm p} = \frac{v_{\bm p}}{u_{\bm p}} \sigma^1 \, , \qquad 
 Y_{\bm p} = \frac{1}{u_{\bm p}} I \, , \label{defXYZ}
\end{equation}
and 
\begin{equation}
 w ({\bm p}) = 
 \left(
\begin{array}{c}
 w_{\bm p} \\
 w_{-{\bm p}}
\end{array}
 \right) \, . 
\end{equation}
In the definition of the matrices $X_{\bm p}$ in Eq.~(\ref{defXYZ}), we used
the Pauli matrix: 
\begin{equation}
 \sigma^1 = 
 \left(
\begin{array}{cc}
 0 & 1 \\
 1 & 0
\end{array}
 \right) \, . 
\end{equation}
The $w$- and $z$-independent normalization factor in Eq. (\ref{tSz})
is fixed by the relation,
\begin{eqnarray}
\exp \left(w^\ast ({\bm p}) w' ({\bm p}) \right) & = & \braket{w({\bm
 p})}{w'({\bm p})} \nn \\ 
& = & \bra{w({\bm p})} U_{\bm p} U_{\bm p}^\dag \ket{w'({\bm p})} \nn \\
& = & \int \! {\rm d} \mu (z({\bm p})) \, \bra{w({\bm p})} U_{\bm p}
 \ket{z({\bm p})} \bra{z({\bm p})} U_{\bm p}^\dag \ket{w'({\bm p})} \nn \\
& = & \int \! {\rm d} \mu (z({\bm p})) \, \bra{w({\bm p})} U_{\bm p}
 \ket{z({\bm p})} \left( \bra{w'({\bm p})} U_{\bm p} \ket{z({\bm p})}
 \right)^\ast  \, , 
\end{eqnarray}
where the formula (\ref{genGaussInt}) can be applied to the last
expression of the integral. 
Thus we obtain an explicit operator representation of $U_{\bm p}$ with the help
of the relation $A(w_{\bm p}^\ast, w_{- {\bm p}}^\ast, z_{\bm p},
z_{-{\bm p}}) = \bra{w({\bm p})} \! :
\! A ({a}_{\bm p}^\dag, a_{-{\bm p}}^\dag, a_{\bm p}, a_{-{\bm p}}) \! :
\! \ket{z({\bm p})} / \braket{w({\bm p})}{z({\bm p})}$ where the symbol
$: \cdots :$ implies the normal ordering to keep the creation operators
to the left of the annihilation operators, 
\begin{equation}
 U_{\bm p} = \frac{1}{u_{\bm p}} : \exp \left[ \frac{1}{2}
 a^{\dag {\rm T}} ({\bm p}) X_{\bm p} {a}^\dag ({\bm p}) - \frac{1}{2}
 {a}^{\rm T} ({\bm p}) X_{\bm p} {a} ({\bm p}) + {a}^{\dag {\rm T}}
 ({\bm p}) Y_{\bm p} {a} ({\bm p}) \right] : \, , 
\end{equation}
where
\begin{equation}
 a ({\bm p}) = 
\left(
\begin{array}{c}
 a_{\bm p} \\
 a_{-{\bm p}}
\end{array}
\right) \, , \qquad
a^\dag  ({\bm p}) = 
\left(
\begin{array}{c}
 a_{\bm p}^\dag \\
 a_{-{\bm p}}^\dag
\end{array}
\right).
\end{equation}

Next, we consider the ${\bm p} = {\bm 0}$ part of $U$. 
The coherent state of the ${\bm p} = {\bm 0}$ mode as 
\begin{equation}
 \ket{z ({\bm 0})} = \exp [ z_{\bm 0} a_{\bm 0}^\dag] \KV = \left[
 \sum_{\ell = 0}^{\infty} \frac{\left( z_{\bm 0} a_{\bm 0}^\dag
 \right)^\ell}{\ell !}\right] \KV \, ,
\end{equation}
where 
\begin{eqnarray}
 a_{\bm 0} \ket{z({\bm 0})} & = & z_{\bm 0} \ket{z ({\bm 0})} \, ,
  \quad 
 a_{\bm 0}^\dag \ket{z({\bm 0})} = \partial_{z_{\bm p}} \ket{z ({\bm
 0})} \, , \\
 \bra{z ({\bm 0})} a_{\bm 0} & = & \partial_{z^\ast_{\bm 0}} \bra{z
  ({\bm 0})} \, , \quad \bra{z ({\bm 0})} a_{\bm 0}^\dag = z_{\bm
  0}^\ast \bra{z ({\bm 0})} \, , 
\end{eqnarray}
and 
\begin{equation}
 \braket{0}{z({\bm 0})} = 1 .
\end{equation}
The completeness condition of the coherent state reads
\begin{equation}
 \int \! {\rm d} \mu (z ({\bm 0})) = \hat{1} \, ,
\end{equation}
where the measure is defined as 
\begin{eqnarray}
 {\rm d} \mu (z ({\bm 0})) & = & e^{- |z_{\bm 0}|^2} \frac{{\rm d} x_{\bm 0} \, {\rm d}
 y_{\bm 0}}{\pi} \, , \\
 z_{\bm 0} & = & x_{\bm 0} + i y_{\bm 0} \, . 
\end{eqnarray}
The following Gaussian integral is relevant here:
\begin{eqnarray}
 && \int \! {\rm d} \mu (z ({\bm 0})) \exp \left[ - \frac{1}{2} \left( M' z_{\bm
 0}^2 + {N'}^\ast z_{\bm 0}^{\ast 2} \right) + u' z_{\bm 0} + v^{'
 \ast} z_{\bm 0}^\ast \right] \nn \\
 & = &  ( 1 - M' N^{' \ast})^{-1} \exp \left[ - \frac{1}{2}
  \frac{{N'}^\ast {u'}^2 + M' {v'}^{\ast 2} - 2 u' {v'}^\ast}{1 - M'
  {N'}^\ast} \right] \, , 
\end{eqnarray}
where $N'$, ${M'}^\ast$, $u'$, ${v'}^\ast$ are constants. 
In a manner similar to the manipulations of
the ${\bm p} \ne {\bm 0}$ part of the $U$, we obtain 
the following equations, 
\begin{eqnarray}
 (u_{\bm 0} \partial_{w_{\bm 0}^\ast} - v_{\bm 0} w_{\bm 0}^\ast -
  z_{\bm 0}) \bra{w ({\bm 0})} U_{\bm 0} \ket{z ({\bm 0})} & = & 0 \, ,
  \\  
 (- v_{\bm 0} \partial_{w_{\bm 0}^\ast} + u_{\bm 0} w_{\bm 0}^\ast -
  \partial_{z_{\bm 0}}) \bra{w ({\bm 0})} U_{\bm 0} \ket{z ({\bm 0})} &
  = & 0 \, .
\end{eqnarray}
One easily find the matrix elements of $U_{\bm 0}$ as 
\begin{equation}
 \bra{w ({\bm 0})} U_{\bm 0} \ket{z ({\bm 0})} = \frac{1}{\sqrt{u_{\bm
  0}}} \exp \left[ \frac{1}{2} \frac{v_{\bm 0}}{u_{\bm 0}} ( w_{\bm
  0}^{\ast 2} - z_{\bm 0}^2 ) + \frac{1}{u_{\bm 0}} w_{\bm 0}^\ast
  z_{\bm 0} \right] \, , 
\end{equation}
so the representation of $U_{\bm 0}$ in terms of $a_{\bm p}$ and $a_{\bm
0}^\dag$ is given as
\begin{equation}
 U_{\bm 0} = \frac{1}{\sqrt{u_{\bm 0}}} :\exp \left[ \frac{1}{2}
\frac{v_{\bm 0}}{u_{\bm 0}} (a_{\bm 0}^{\dag 2} - a_{\bm 0}^2 ) +
\frac{1}{u_{\bm 0}} a_{\bm 0}^\dag a_{\bm 0} \right]: \, . 
\end{equation}

As a summary of this appendix,
the vacuum of the quasi-particle, described by the $b_{\bm p}$
operators including the ${\bm p} = {\bm 0}$ case, is explicitly given by  
\begin{eqnarray}
  \ket{\Omega_0} & = &  U \KV \nn \\
 & = & \left[ \frac{1}{\sqrt{u_{\bm 0}}} \exp \left\{ \frac{1}{2}
 \frac{v_{\bm 0}}{u_{\bm 0}} a_{\bm 0}^{\dag 2} \right\} \right] \left[
 {\prod_{{\bm p} = {\bm 0}}^\infty}' \frac{1}{u_{\bm p}}  
 \exp \left\{ \frac{1}{2} {a}^{\dag {\rm T}} ({\bm p}) X_{\bm p} {a}^\dag
 ({\bm p}) \right\} \right] \KV \nn \\ 
 & = & \frac{1}{\sqrt{u_{\bm 0}}} \exp \left[ - \frac{1}{2} {\sum_{{\bm p} = {\bm
 -\infty}}^{\infty}}' \ln u_{\bm p} \right] \exp \left[ \frac{1}{2}\sum_{{\bm p} = {\bm
 -\infty}}^{\infty} \frac{v_{\bm p}}{u_{\bm p}} a_{\bm p}^\dag a_{-{\bm
 p}}^\dag \right]  \KV \, ,  \label{S0} 
\end{eqnarray}
where ${\prod_{{\bm p} = -\infty}^\infty}'$ and 
${\sum_{{\bm p} = {\bm 0}}^{\infty}}'$ mean product and summation 
without ${\bm p} = {\bm 0}$, respectively.

\end{document}